\documentclass[journal]{IEEEtran}
\ifCLASSINFOpdf
   \usepackage[pdftex]{graphicx}

\else

\fi

\usepackage{xcolor}
\usepackage{layouts}
\usepackage{mathtools}
\usepackage{cuted}
\usepackage{booktabs}
\usepackage{amsfonts}
\usepackage{amssymb}
\usepackage{amsmath}
\usepackage{cite}
\usepackage{amssymb, amsmath, amsthm, amsfonts}
\usepackage{caption}
\usepackage{subcaption}
\newcommand{\bx}{\mathbf{x}}
\newcommand{\btheta}{\boldsymbol{\theta}}
\newcommand{\bmu}{\boldsymbol{\mu}}
\newcommand{\bV}{\mathbf{V}}

\newcommand{\init}{\text{init}}

\newcommand{\st}{\text{s.t.}}

\newcommand{\R}{\mathbb{R}}

\newtheorem{theorem}{Theorem}

\newtheorem{definition}{Definition}

\begin{document}
\title{An Iterative Approach to Improving Solution Quality for AC Optimal Power Flow Problems
}

\author{Ling Zhang,~\IEEEmembership{Student Member,~IEEE,}
        and~Baosen~Zhang,~\IEEEmembership{Member,~IEEE,}%
\thanks{L. Zhang and B. Zhang are with the Department
of Electrical and Computer Engineering, University of Washington, WA, 98195 USA e-mail: \{lzhang18,zhangbao\}@uw.edu 
The authors are partially supported by NSF grants ECCS-1942326 and ECCS-2023531}
}

\maketitle

\begin{abstract}
The existence of multiple solutions to AC optimal power flow (ACOPF) problems has been noted for decades. Existing solvers are generally successful in finding local solutions, which satisfy first and second order optimality conditions, but may not be globally optimal. In this paper, we propose a simple iterative approach to improve the quality of solutions to ACOPF problems. 
First, we call an existing solver for the ACOPF problem. From the solution and the associated dual variables, we form a partial Lagrangian. Then we optimize this partial Lagrangian and use its solution as a warm start to call the solver again for the ACOPF problem. 
By repeating this process, we can iteratively improve the solution quality, moving from local solutions to global ones.
We show the effectiveness our algorithm on standard IEEE networks. The simulation results show that our algorithm can escape from local solutions to achieve global optimums within a few iterations.

\end{abstract}


\IEEEpeerreviewmaketitle

\section{Introduction}

The optimal power flow (OPF) problem is a fundamental resource allocation problem in power system operations that minimizes the cost of power generation while satisfying demand. The ACOPF formulation of the problem uses nonlinear power flow equations, resulting in nonlinear and nonconvex optimization problems~\cite{cain2012history,Molzahn19,hiskens2001exploring}.  
The consequence of the nonconvexity of ACOPF we study in this paper is the presence of multiple solutions. 

Most ACOPF problems are solved via variations of nonlinear optimization algorithms, including Newton-Raphson, sequential programming, interior points and others (see~\cite{Molzahn19,qiu2009literature,capitanescu2016critical} and the references within). These algorithms are in general only able to certify a solution is locally optimal, that is, they satisfy first order and/or second order optimality conditions. Because the existence of many such local solutions, it is often difficult to find the global optimal one and operate the system at with the least cost. 

The existence of multiple local solutions of the OPF problem has been well-known for several decades~\cite{ma1993efficient,momoh1999review,lesieutre2015efficient}. Despite this, a common assumption is that OPF problems tend to have a single ``practical'' solution that is globally optimal, and therefore the fact that multiple solutions can exist do not impact day-to-day operations~\cite{momoh1997challenges,Wei98}. However, an increasingly large body of work have pointed to that multiple solutions do occur under reasonable conditions and cannot easily ruled out~\cite{bukhsh2013local,wu2017deterministic,Molzahn19}. For example, \cite{bukhsh2013local} shows how modifications of the standard IEEE benchmarks can lead to each having more than one local solutions. Statistical studies in \cite{lesieutre2019distribution,lindberg2020distribution} show that there are more solutions than previously thought in many systems. 

An open question in the field is to develop algorithms that can find global optimal solutions, or at least improve upon local ones. In addition to lowering the operating cost, understanding and distinguishing between local and global optimal solutions can lead to important theoretical discoveries about the ACOPF problem. Consequently, several classes of algorithms have been developed. For example, holomorphic embedding has been used in~\cite{li2021implications, dronamraju2021implications}, but are slow and require very high numerical precision. Genetic algorithms can escape a local minimum, but are random in nature and require repeated trial and error~\cite{bakirtzis2002optimal,abido2002optimal}. Robust optimal power flow can alleviate convergence issues, but may not improve on the quality of the solution~\cite{oh2019unified}.  

In this paper, we propose a simple algorithm that can effectively escape from strict local solutions to find better ones. That is, we can move from one solution to another while reducing the objective value, and may therefore move towards the global optimal solutions. This algorithm is deterministic, relies on duality theory and uses existing solvers as subroutines. 

\begin{figure}[ht]
    \centering
    \includegraphics[scale=0.6]{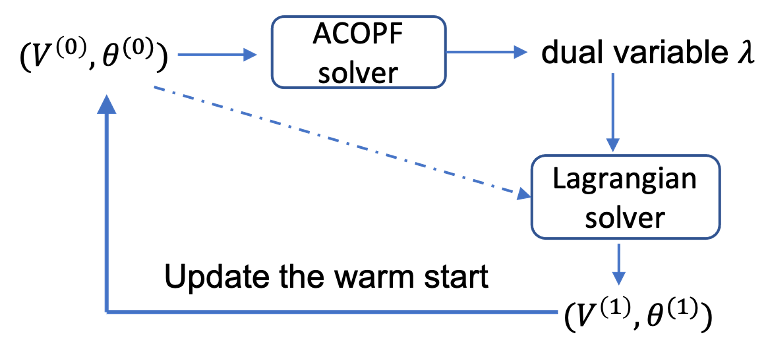}
    \caption{Outline of our algorithm. We form partial Lagragians at local optima and solve it to provide better warm starting points to an ACOPF solver.}
    \label{fig:intro}
\end{figure}

Our process is outlined in Fig.~\ref{fig:intro}. First, we solve the ACOPF problem using an existing solver (e.g., IPOPT~\cite{wachter2006implementation} or Matpower~\cite{zimmerman2010matpower}). From the solution and its associated dual variables, we form a partial Lagrangian. This partial Lagrangian serves to reshape the geometry of the optimization problem. We then optimize this partial Lagrangian, which can lead to a different solution. Using this second solution as a warm start, we again call the solver for the ACOPF problem. Repeating this iterative process, we can successively improve the solution quality, moving from higher cost solutions to lower cost ones. 

We provide both theoretical analysis on tree networks and simulation results on (meshed) 3-bus 9-bus, 22-bus and 39-bus networks.
We show that our algorithm can quickly escape from local solutions and find lower cost solutions. This feature holds even for ACOPF problems with disconnected feasible spaces, which has been traditionally difficult to deal with~\cite{Molzahn19,bukhsh2013local}. For networks with known global solutions (3, 9, 22-bus), we show that our algorithm can find the global optimal solution in a single iteration, even starting from a strictly local solution.

Our approach can be seen as a way to provide good warm starts to nonlinear optimization solvers. This is commonly done using the linearized DCOPF, although it can fail to find good starting points as shown by our simulations as well as existing results~\cite{baker2021solutions}. More sophisticated approaches either randomizes~\cite{CE21} or uses a previous solution as the starting point~\cite{tang2017distributed}. The former tends to be time consuming, while the latter tend lead to system being stuck in a strict local solution~\cite{bukhsh2013local}. The work in \cite{mulvaney2020load} suggests that solutions can escape local minima if load undergoes random fluctuations. Our approach can be seen as providing an explicit and deterministic algorithm to search for global solutions by using dual variables at local solutions. 


\section{Model and Problem Formulation} \label{sec:model}
Consider a power system network where $n$ buses are connected by $m$ edges. 
For bus $i$, let $V_i$ denote its voltage magnitude, $\theta_i$ its angle, $P^{G}_i$ and $Q^{G}_i$ the active and reactive output of the generator and $P^{D}_i$ and $Q^{D}_i$ the active and reactive load. We use $P^{f}_{ij}$ and $Q^{f}_{ij}$ to denote the active and reactive power flowing from bus $i$ to bus $j$. The admittance between buses $i$ and $j$ is $g_{ij}-b_{ij}$. 
We use $\theta_{ij}$ as a shorthand for $\theta_i-\theta_j$. 

The ACOPF problem is to minimize the cost of active power generations while satisfying a set of constraints~\cite{bukhsh2013local}:
\begin{subequations} \label{prob1}
\begin{align}
     \min_{\bV, \btheta} &\textstyle \sum_{i} c_i(P^{G}_i)\\
    \st ~ & P^{G}_i = P^{D}_i + \textstyle \sum_{j=1}^{N} P^{f}_{ij}\label{Pbalanc}\\
    & Q^{G}_i = Q^{D}_i + \textstyle \sum_{j=1}^{N} Q^{f}_{ij}\label{Qbalanc}\\
    & P^{f}_{ij} = V_i^2g_{ij}-V_iV_j(g_{ij}\cos(\theta_{ij})-b_{ij}\sin(\theta_{ij}))\label{PEq}\\
    & Q^{f}_{ij} = V_i^2 \hat{b}_{ij} - V_iV_j(b_{ij}\cos(\theta_{ij}) +g_{ij}\sin(\theta_{ij}))\label{QEq}\\
    & \underbar{V}_i\leq V_i \leq \bar{V}_i\label{Vlimits}\\
    & \underbar{P}^{G}_i\leq P^G_i \leq \bar{P}^{G}_i\label{PGlimits}\\
    & \underbar{Q}^{G}_i\leq Q^G_i \leq\bar{Q}^{G}_i\label{QGlimits}\\
    & (P^{f}_{ij})^2+(Q^{f}_{ij})^2\leq (S_{ij}^{\max})^2\label{flimits}
\end{align}
\end{subequations}
where $\hat{b}_{ij}=b_{ij}-0.5b_{ij}^C$ and $b_{ij}^C$ is the line charging susceptance. The constraints  (\ref{Pbalanc}) and (\ref{Qbalanc}) enforce power balance, (\ref{PEq}) and (\ref{QEq}) are the AC power flow equations, (\ref{Vlimits}) limits the bus voltage magnitudes, (\ref{PGlimits}) and (\ref{QGlimits}) represent the active and reactive limits and \eqref{flimits} are the line flow limits. We assume the cost at each bus $i$, $c_i(\cdot)$, is increasing. Other than that, the cost can be linear, quadratic or other functions. 


We assume problem \eqref{prob1} is feasible in this paper. Out of the feasible solutions, we focus on two classes: local solutions and global solutions. Local solutions are all the solutions that satisfies local optimally conditions, for example, the KKT conditions or second order ones~\cite{bertsekas1997nonlinear}. Out of this set, the solutions with the lowest cost are called the global ones. We sometimes refer to the local solutions that are not global as strict local solutions. 

Over the years, many nonlinear programming (NLP) solvers have been developed for the ACOPF problem, and their speed and efficiency have improved dramatically (e.g.,~see~\cite{cain2012history} and the references within). However, NLP solvers are typically only able to return local solutions. Since a local solution is not necessarily global, we propose an iterative approach to improve the solution quality by alternatively solving (\ref{prob1}) and a partial Lagrangian. Any NLP solver can be used, and and we use IPOPT \cite{wachter2006implementation} in this paper.

\section{Algorithm} \label{sec:algo}
Our algorithm starts with a call to a NLP solver with some initial guess, denoted by $\btheta_{\init}$, $\bV_{\init}$. For example, this can be the standard flat start with voltage magnitudes being 1 p.u. and angles set to $0$. 
Then we assume the solver returns a feasible solution. Of course, we don't know whether this solution is globally optimal. At this solution, we record the dual variables associated with the power balance equations \eqref{Pbalanc} and \eqref{Qbalanc}, denoted as $\bar{\bmu}^P$ and $\bar{\bmu}^Q$. Using these dual variables, we form the following partial Lagrangian by dualizing the power balance equations: 
\begin{subequations}
\begin{align}
\mathcal{L}(\bV, \btheta, \bmu^P, \bmu^Q)=&  \sum_{i} c_iP^{G}_i+ \sum_{i}\mu_i^{P}(P^{D}_i + \sum_{j=1}^{N} P^{f}_{ij} - P^{G}_i)\nonumber\\
    & +  \sum_{i}\mu_i^Q(Q^{D}_i + \sum_{j=1}^{N} Q^{f}_{ij} - Q^{G}_i). \nonumber 
\end{align}
\end{subequations}
We then minimize the partial Lagrangian by solving 
\begin{align}
  \min_{\bV, \btheta} \; & \mathcal{L}(\bV, \btheta, \bmu^P, \bmu^Q) \label{prob2} \\
    \st & ~ (\ref{PEq})- (\ref{flimits}). \nonumber
\end{align}
The problem in~\eqref{prob2} can be solved using any NLP solver, and is feasible if the original ACOPF problem is feasible. 

We solve the problem in \eqref{prob2}
\emph{starting from the same initial point $(\bV_{\init}, \btheta_{\init})$} that was used to solve the original primal problem in \eqref{prob1}. Denote this solution to \eqref{prob2} by $(\bar{\bV}, \bar{\btheta})$. Note $(\bar{\bV}, \bar{\btheta})$ will not be the same as $(\bV_{\init}, \btheta_{\init})$ since they come from different problems. Then we start the NLP solver again to solve (\ref{prob1}) but with the initial point $(\bar{\bV}, \bar{\btheta})$. This process can be repeated until the solutions stop changing or up to a predefined number of iterations.

It turns out the solution $(\bar{\bV}, \bar{\btheta})$ found by solving the partial Lagrangian is often a much better starting point than the original choice of $(\bV_{\init}, \btheta_{\init})$. Therefore, by repeating these steps, we can iteratively improve the solution quality (i.e., reducing the cost). The algorithm is summarized below as Algorithm 1. We illustrate the intuition behind this algorithm in the next section using 2-bus and 3-bus networks. Formal proofs are given in Section~\ref{sec:analysis}, and simulations results for larger IEEE benchmarks are presented in Section~\ref{sec:results}. 
\begin{table}[ht]
\normalsize
\begin{tabular}{ll}
\hline
\multicolumn{2}{l}{\textbf{Algorithm 1:  Solving ACOPF iteratively}}\\
\hline
\multicolumn{2}{l}{\textbf{Inputs:}~$\btheta_{\init}^{(i)}$, $\bV_{\init}^{(i)}$, $i=0$}\\
1:~At $i$-th iteration: Initialized at $\btheta_{\init}^{(i)}$, $\bV_{\init}^{(i)}$:\\
2:~Call NLP solver for \eqref{prob1}, record $(\bar{\bmu}^P_{(i)}, \bar{\bmu}^Q_{(i)})$.\\
3:~Given $(\bar{\bmu}^P_{(i)}, \bar{\bmu}^Q_{(i)})$, call solver the partial \\
~~~Lagrangian in \eqref{prob2}, record the solutions as $(\bar{\btheta}^{(i)}, \bar{\bV}^{(i)})$.\\
4:~Call IPOPT for \eqref{prob1} initialized at $(\bar{\btheta}^{(i)}$, $\bar{\bV}^{(i)})$, \\
~~~record solutions $(\hat{\btheta}^{(i)}, \hat{\bV}^{(i)})$.\\
5:~If the solution from line 4 does not reduce the cost, \\
~~~terminate the algorithm.\\
6:~Otherwise, update initial points:\\
~~~$\btheta_{\init}^{(i+1)} = \hat{\btheta}^{(i)}$, $\bV_{\init}^{(i+1)} = \hat{\bV}^{(i)}$.\\
7:~Repeat the until the maximum number of \\
~~~iterations is reached.\\
\hline
\end{tabular}
\label{algo}
\end{table}

In terms of computational overhead,  each iteration of Algorithm 1 solves an ACOPF problem twice and an OPF-like problem (minimizing the partial Lagrangian) once. In practice, we observe that the cost is reduced after every iteration and the global solution can be reached in a small number of iterations (for the cases where the global iteration is known). Therefore, in contrast to algorithms that resolve the ACOPF problem from a large number of random initialization points \cite{CE21}, Algorithm 1 is much more computationally efficient.

\vspace{-0.5cm}
\section{Geometry and Intuition}
\label{sec:geometry}
In this section, we study the geometry of the ACOPF problem to shed some light on why Algorithm 1 might be successful. 
We find that the main reason is that the optmization landscape of the partial Lagrangian is much ``better'' than the landscape of the original problem. 
To illustrate this geometric property, we
use the 2-bus and 3-bus networks as examples. The formal proofs are provided in Section~\ref{sec:analysis}.

\subsection{2-bus network}\label{sec:geometry2bus}
In this part, we consider a 2-bus network. 
For simplicity, we ignore the reactive power and set both voltage magnitudes to 1 p.u.. 
Suppose bus 1 is a generator and the reference (slack) bus with an increasing cost function $c(\cdot)$, and bus 2 is the load bus with angle $-\theta$. The line admittance is $g-jb$. 
Given a load of $l$ at bus 2 and ignoring all constraints except for the load balancing one, the ACOPF in \eqref{prob1} becomes
\begin{subequations} \label{eqn:two_bus}
\begin{align}
    \min_{\theta} \; & c(g-g\cos(\theta)+b\sin(\theta))\\
    \st~& l+g-g\cos(\theta)-b\sin(\theta)=0\label{eq:2bus}. 
\end{align}
\end{subequations}
This is an example of an OPF with a disconnected feasible space, since there are two discrete solutions to \eqref{eq:2bus} and we are asking for the lower cost one. 

\begin{figure}[t]
\centering
\includegraphics[height=5cm, width=6cm]{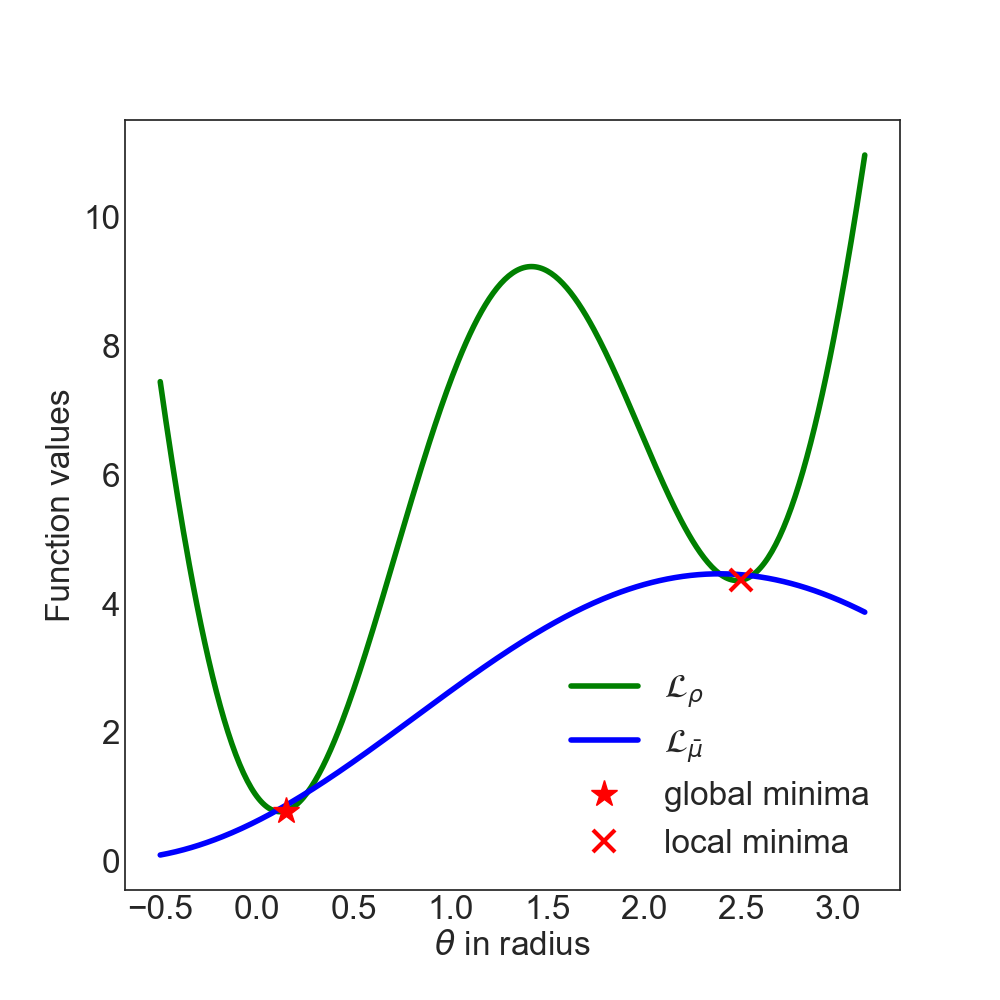}
\caption{Geometry of the penalized objective functions $\mathcal{L}_{\rho}$ and the partial Lagrangian $\mathcal{L}_{\mu}$. The line admittance is $1-j4$ and the penalty parameter is $2$.
\vspace{-0.5cm}}
\label{fig:2bus}
\end{figure}

To see how a NLP solver would approach this problem, we adopt a common practice~\cite{bertsekas1997nonlinear,mulvaney2020load} and form a penalized version of~\eqref{eqn:two_bus}. The penalized unconstrained problem is given by 
\begin{align}\label{eqn:pentwobus}
    \mathcal{L}_{\rho}= & c(g-g\cos(\theta)+b\sin(\theta)) \\ 
    &+\rho/2(l+g-g\cos(\theta)-b\sin(\theta))^2, \nonumber
\end{align}
where $\rho$ is a penalty parameter. For large enough $\rho$, the solutions of \eqref{eqn:pentwobus} would coincide with those of~\eqref{eqn:two_bus}~\cite{bertsekas1997nonlinear}. The function $\mathcal{L}_{\rho}$ is plotted in Fig.~\ref{fig:2bus} (green line). We can see that there are two local minima, with the left one being global. The strict local minimum (the right one) satisfies both first and second order optimality conditions. Therefore, if we initialize a NLP solver with a bad starting point, it be would stuck at the strict local solution. For this example, if the initial point is to the left of the maximum of the green curve, a solver would converge to the left solution; and if the initial point is to the right, a solver would find the right (suboptimal) solution. Hence, a flat start would lead to the global solution. However, for larger systems, flat starts are often not successful (e.g., see the 22-bus system in Section~\ref{sec:results}). Therefore, this 2-bus example is useful as it illustrates the geometry of the optimization landscape. 

Now suppose $\mu$ is the multiplier corresponding to the equality constraint (\ref{eq:2bus}) at \emph{the strict local solution}. The partial Lagrangian of \eqref{eqn:two_bus} by dualizing \eqref{eq:2bus} is:
\begin{align} \label{eqn:Lagrangian2bus}
    \mathcal{L}_{\mu}= & c(g-g\cos(\theta)+b\sin(\theta)) \\
    & +\mu(l+g-g\cos(\theta)-b\sin(\theta)). \nonumber
\end{align}
Since the sinusoidal functions are periodic with period $2\pi$, let us consider the range $\theta\in[-\pi, \pi]$. It is interesting now to compare the solution of $\mathcal{L}_{\mu}$ and the original problem in~\eqref{eqn:two_bus} (or equivalently, $\mathcal{L}_\rho$). The blue curve in Fig.~\ref{fig:2bus} plots $\mathcal{L}_{\mu}$. We observe two interesting facts. The first is that unlike $\mathcal{L}_{\rho}$, $\mathcal{L}_{\mu}$ only has a single minimum. Therefore, no matter where we initialize the NLP solver for $\mathcal{L}_\mu$, we would reach this minimum. The second fact is that the minimum of $\mathcal{L}_\mu$ is close to the global minimum of $\mathcal{L}_\rho$. Therefore, if we start a NLP solver for the ACOPF at the solution of $\mathcal{L}_\mu$, we would reach the global solution. Interestingly, we are using the multiplier at the strict local solution. So even if a solution is not global, it is still very useful, since by solving $\mathcal{L}_\mu$ as an intermediate step, we would not be stuck at the strict local solution. We prove that this procedure is guaranteed to work for tree networks in the next section. 


\captionsetup[figure]{font=small,skip=2pt}



\subsection{3-bus network}\label{sec:three_bus}
Now, let us consider a 3-bus network to show that the intuitions built in the 2-bus example still carryover. We again ignore the reactive power and set all voltage magnitudes to 1 p.u. to optimize over the angles. 
Suppose bus 1 is a generator and also the reference bus with an increasing cost function $c(\cdot)$, while bus 2 and bus 3 are load buses with angles $-\theta_2$ and $-\theta_3$, respectively. 
The load at bus 2 is $l_2$ and at bus 3 is $l_3$. Then the ACOPF in \eqref{prob1} can be simplified to
\begin{subequations}\label{eqn:three_bus}
\begin{align} 
&\min_{\theta_2,\theta_3} c(\sum_{j=2,3} g_{1j}-g_{1j}\cos(\theta_j)+b_{1j}\sin(\theta_j))\\
&\st \nonumber\\
& l_{2}+\sum_{j=1,3}( g_{2j}-g_{2j}\cos(\theta_{2j})-b_{2j}\sin(\theta_{2j})) = 0 \label{eqn:3busb1}\\
& l_{3}+ \sum_{j=1,2}( g_{3j}-g_{3j}\cos(\theta_{3j})-b_{3j}\sin(\theta_{3j}))=0.\label{eqn:3busb2}
\end{align}
\end{subequations}

\begin{figure}[t]
     \centering
    \begin{subfigure}{0.47\columnwidth}
         \centering
         \includegraphics[height=4.2cm, width=4.2cm]{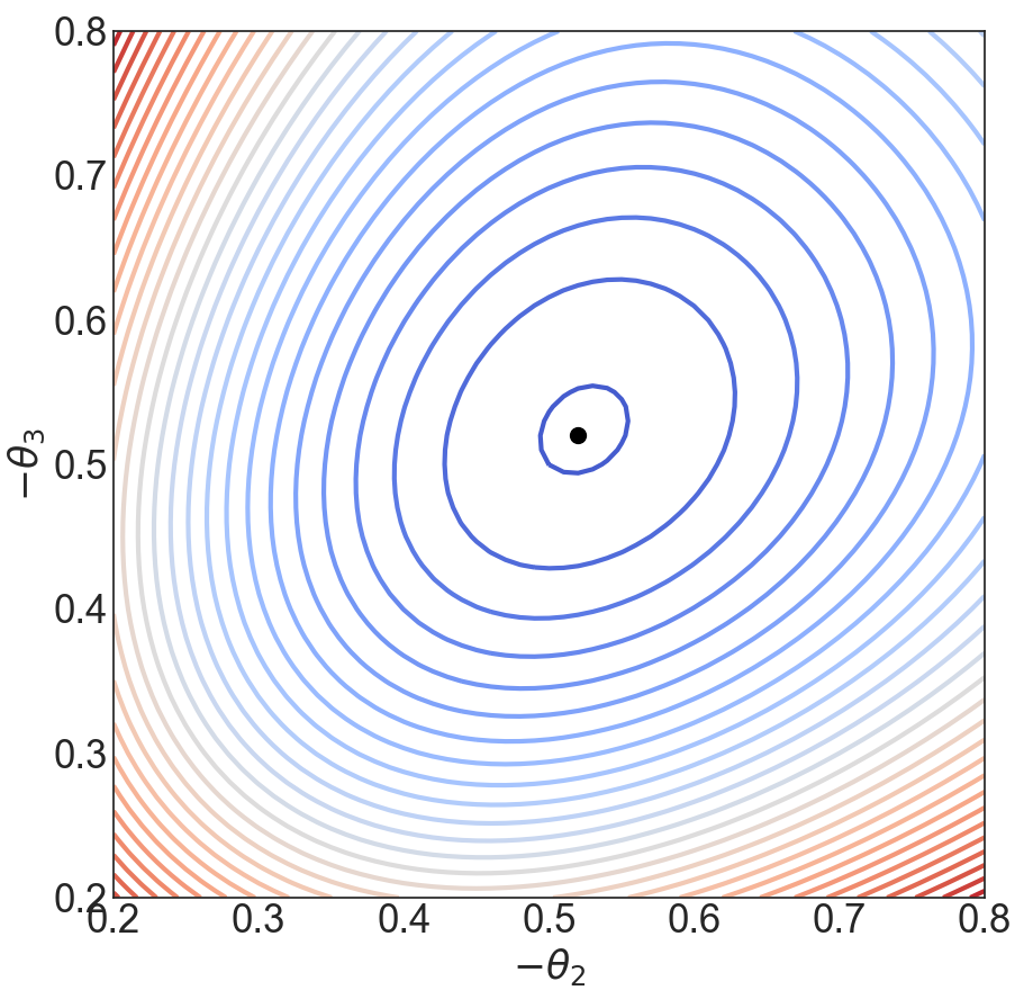}
         \caption{$\mathcal{L}_{\rho}$, global solution}
         \label{fig:y equals x}
     \end{subfigure}
     \begin{subfigure}{0.4\columnwidth}
         \centering
         \includegraphics[height=4.2cm, width=4.2cm]{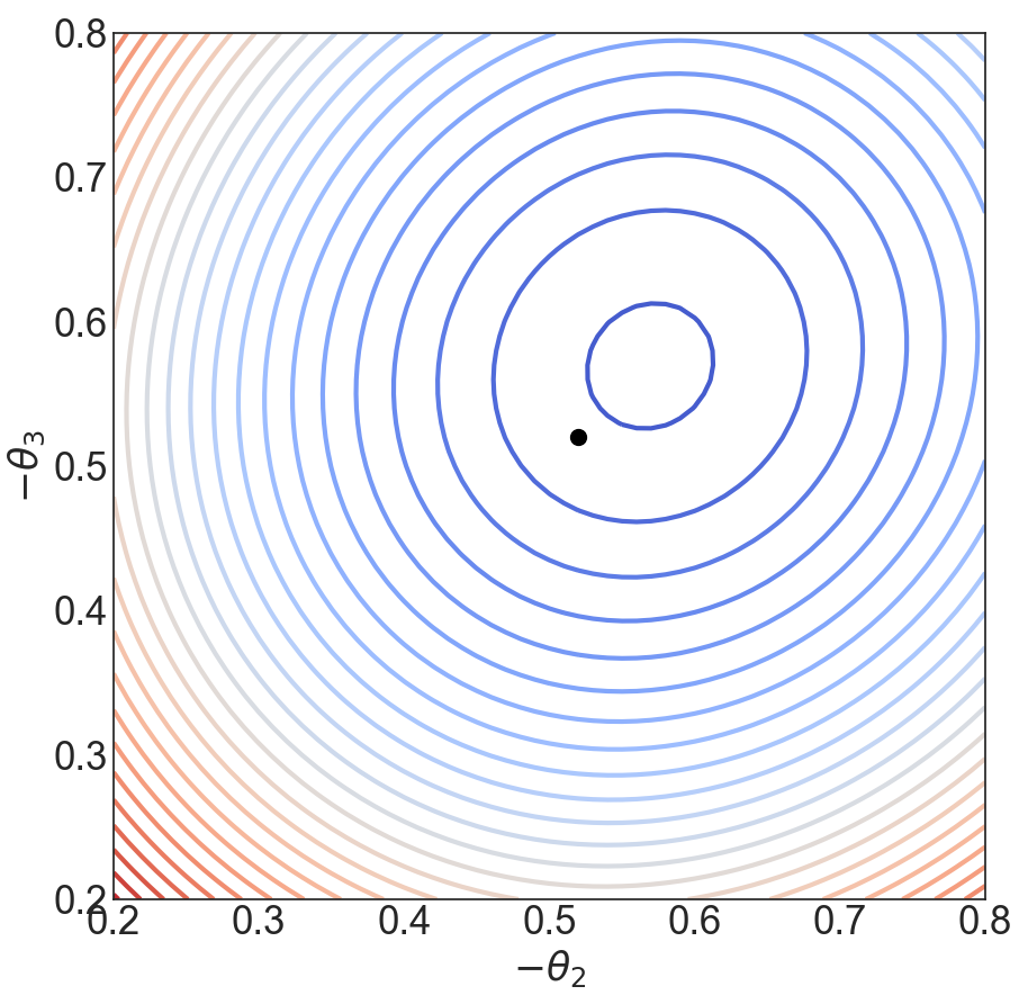}
         \caption{$\mathcal{L}_{\mu}$, global solution}
         \label{fig:three sin x}
     \end{subfigure}
     \begin{subfigure}{0.47\columnwidth}
         \centering
         \includegraphics[height=4.2cm, width=4.2cm]{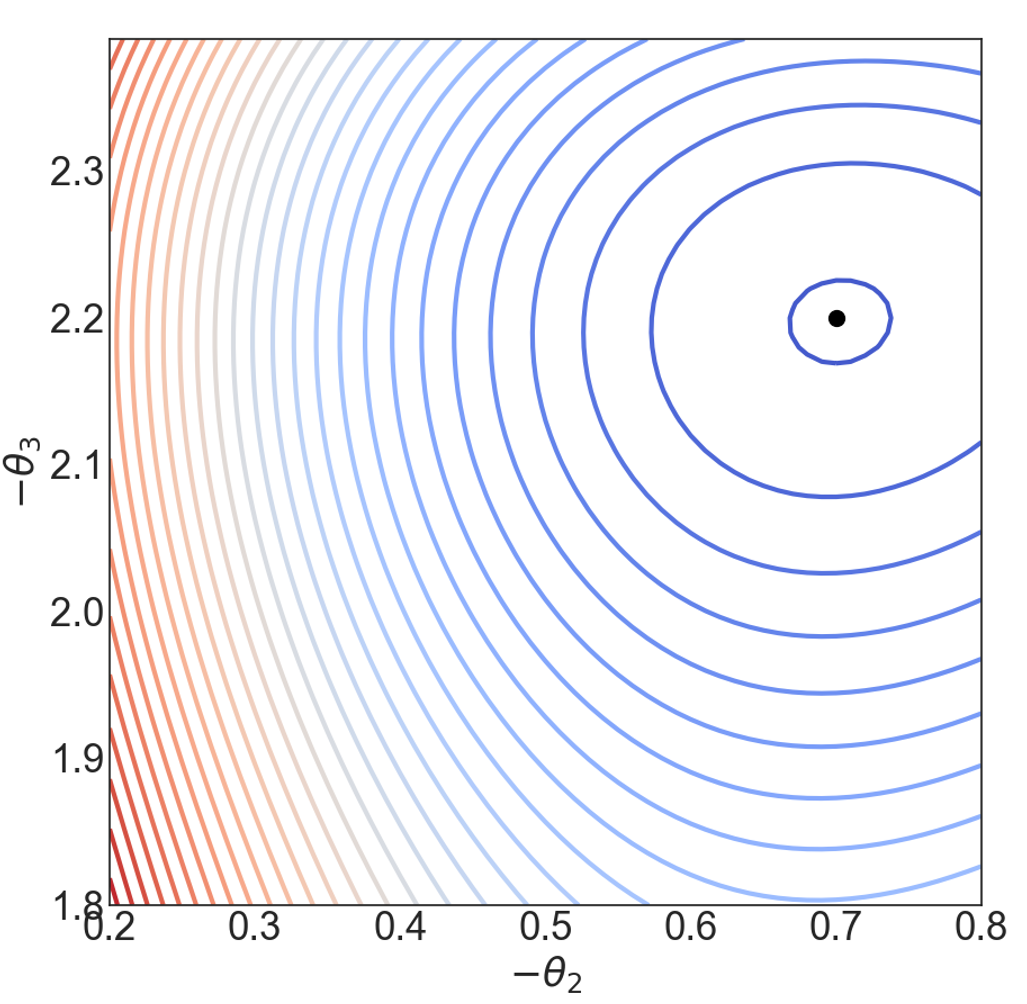}
         \caption{$\mathcal{L}_{\rho}$, local solution}
         \label{fig:y equals x2}
     \end{subfigure}
     \begin{subfigure}{0.4\columnwidth}
         \centering
         \includegraphics[height=4.2cm, width=4.2cm]{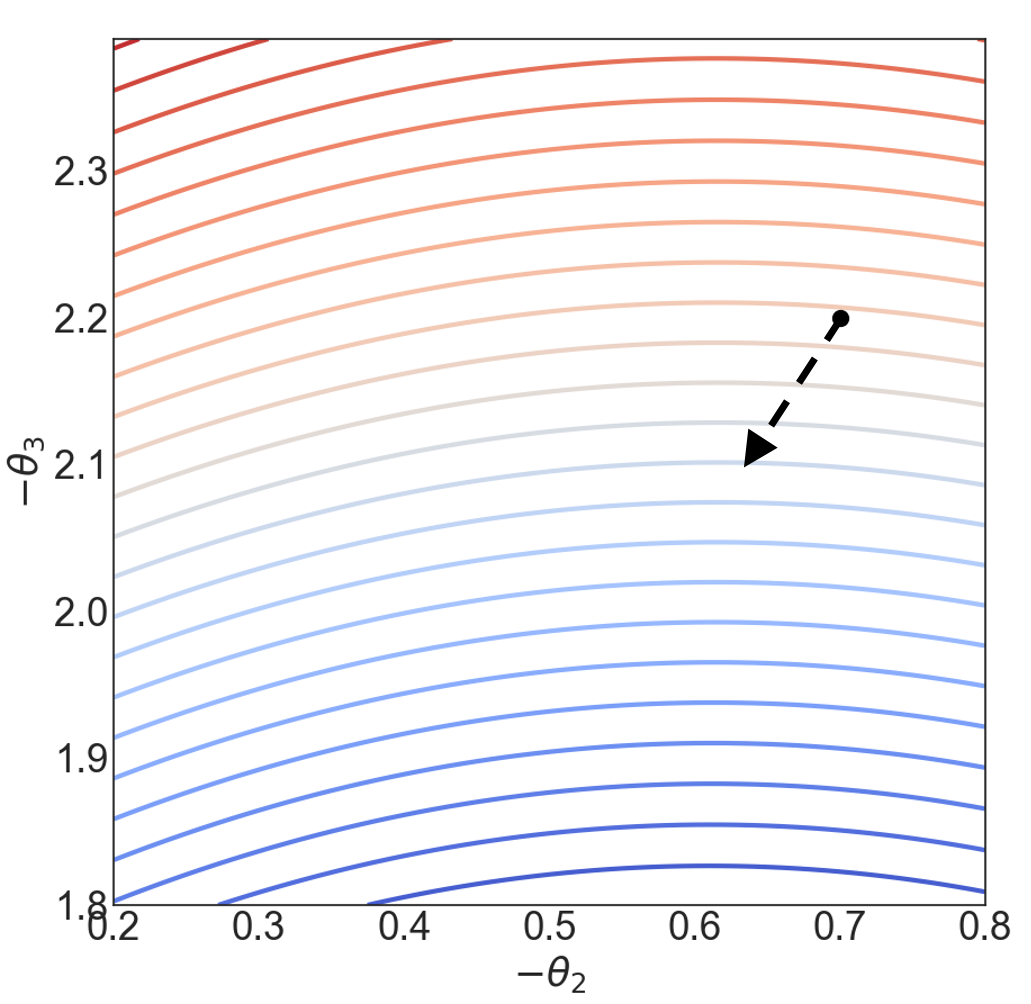}
         \caption{$\mathcal{L}_{\mu}$, local solution}
         \label{fig:three sin x2}
     \end{subfigure}
          \begin{subfigure}{0.47\columnwidth}
         \centering
         \includegraphics[height=4.2cm, width=4.2cm]{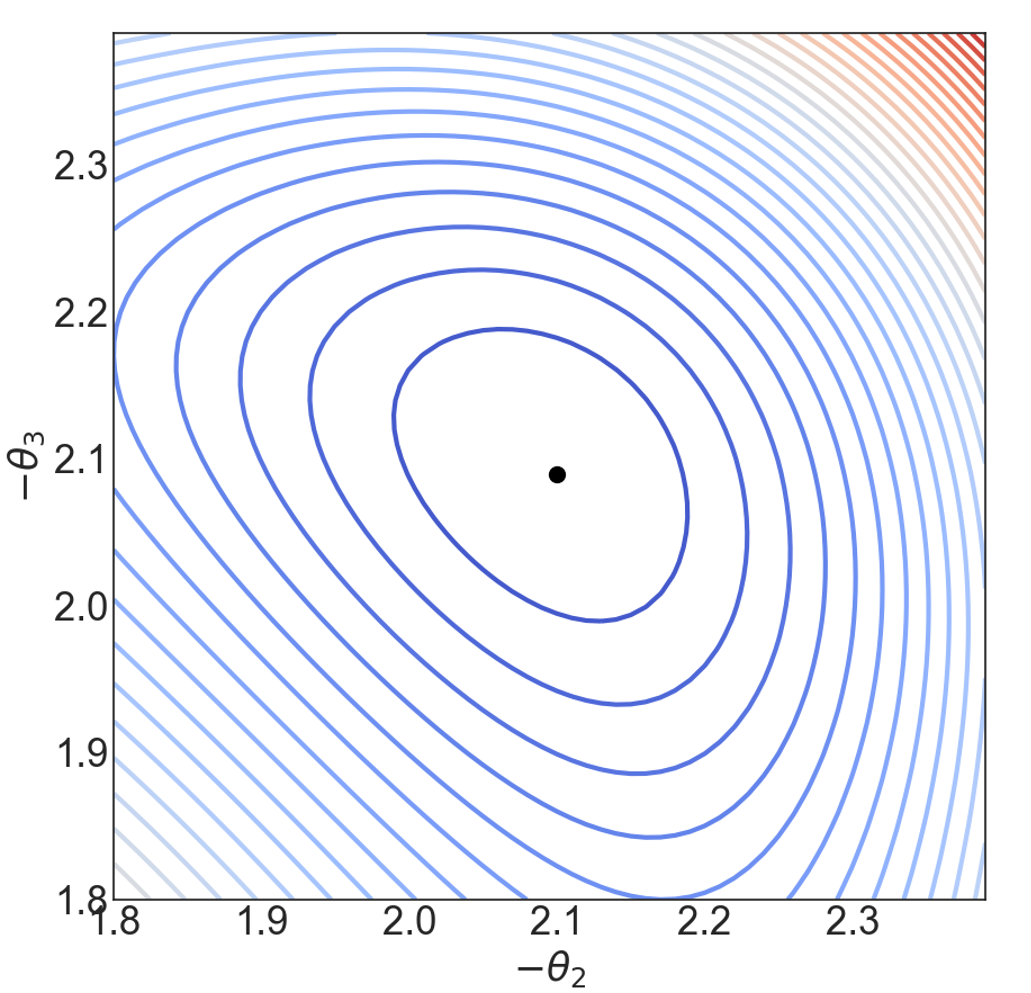}
         \caption{$\mathcal{L}_{\rho}$, local solution}
         \label{fig:y equals x3}
     \end{subfigure}
     \begin{subfigure}{0.4\columnwidth}
         \centering
         \includegraphics[height=4.2cm, width=4.2cm]{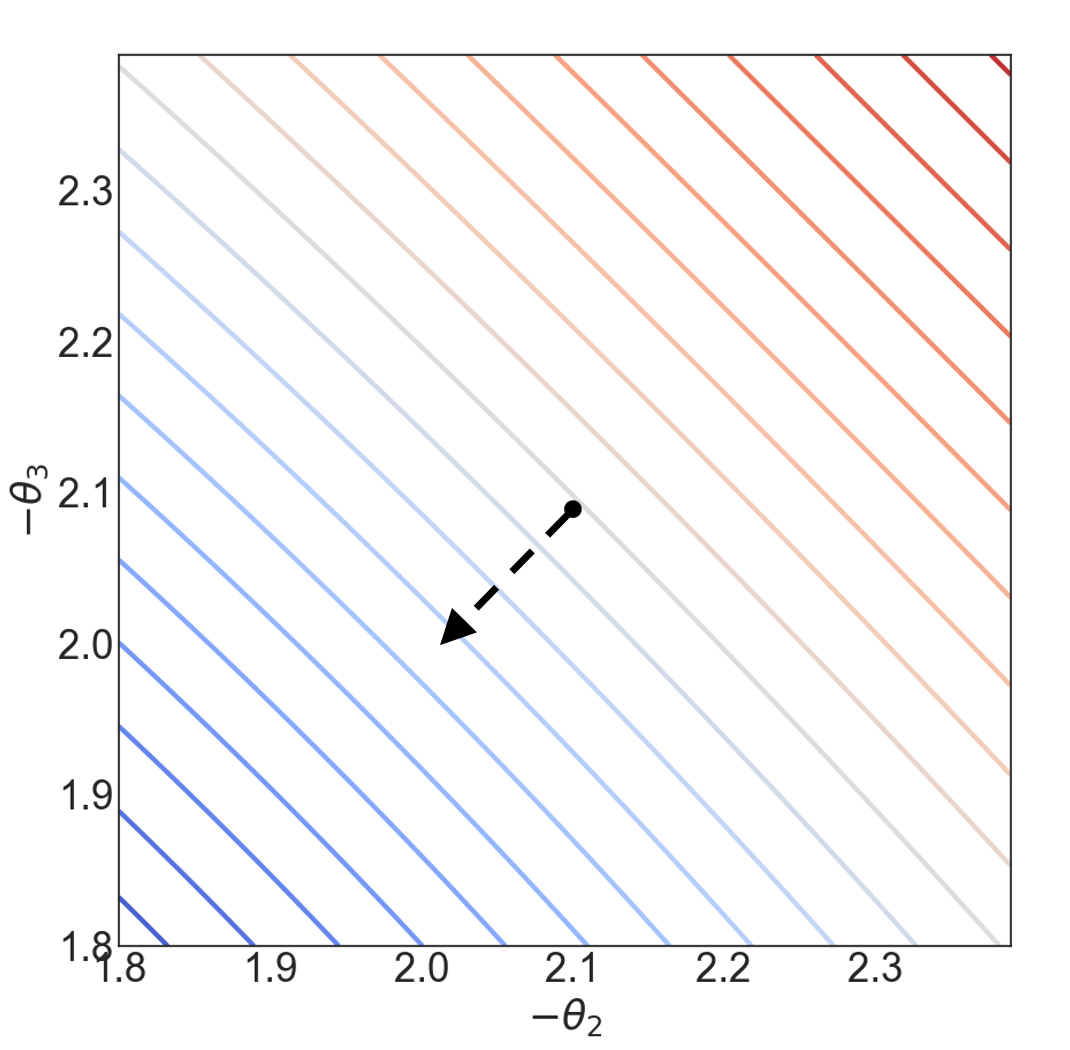}
         \caption{$\mathcal{L}_{\mu}$, local solution}
         \label{fig:three sin x3}
     \end{subfigure}
        \caption{The contour plot of $\mathcal{L}_{\rho}$ and $\mathcal{L}_{\mu}$ nearby the 1st, 2nd and 4th solution. The Hessian matrix of $\mathcal{L}_{\mu}$ is positive definite in (b), indefinite in (d), and negative definite in (f). The black arrows in (d) and (f) indicates the descent directions of the function value. 
        \vspace{-0.5cm}}
        \label{fig:3busHessian}
\end{figure}

As in the 2-bus case, to understand how a NLP solver may approach \eqref{eqn:three_bus}, we form its penalized version:
\begin{align} \label{eqn:penalized3bus}
    &\mathcal{L}_{\rho}=  c(\sum_{j=2,3} g_{1j}-g_{1j}\cos(\theta_j)+b_{1j}\sin(\theta_j)) \\
    &+  \frac{\rho}{2}\left( l_{2}+\sum_{j=1,3}( g_{2j}-g_{2j}\cos(\theta_{2j})-b_{2j}\sin(\theta_{2j})) \right)^2, \nonumber\\
    &+  \frac{\rho}{2}\left( l_{3}+ \sum_{j=1,2}( g_{3j}-g_{3j}\cos(\theta_{3j})-b_{3j}\sin(\theta_{3j})) \right)^2.
    \nonumber
\end{align}
It turns out that there are four local solutions (one of which is global) for \eqref{eqn:penalized3bus}.\footnote{They are found via a grid search.} All of these solutions satisfy both first order and second order conditions are listed in Table~\ref{tab:solns3busA}. At these solutions, gradients $\nabla \mathcal{L}_\rho$ are 0 and the Hessian's $\nabla^2 \mathcal{L}_\rho$ are positive definite. This makes $\mathcal{L}_{\rho}$ look like valleys (convex) at all of the minima and hard for a NLP solver to get out of being trapped at a strict local minima. The level sets around three solutions of $\mathcal{L}_\rho$ are plotted on the left of Fig.~\ref{fig:3busHessian}. They show that there is little difference between the local and the global minima. 

\begin{table}[ht]
\captionsetup{font=small}
\normalsize
\begin{tabular}{llll}
\hline
Solution & Bus 2 & Bus 3 & Hessian matrix of $\mathcal{L}_{\mu}$\\
\hline
1st (global) & $\angle{0.52}$ & $\angle{0.52}$ & Positive definite\\
2nd & $\angle{0.7}$ & $\angle{2.2}$ & Indefinite\\
3rd & $\angle{2.2}$ & $\angle{0.7}$ & Indefinite\\
4th & $\angle{2.09}$ & $\angle{2.09}$ & Negative definite\\
\hline
\end{tabular}
\caption{The four solutions to problem \eqref{eqn:three_bus} through grid search. The Hessian of $\mathcal{L}_{\rho}$ is positive definite at all the solutions. The definiteness of the Hessian of $\mathcal{L}_{\mu}$ is listed. The parameters are $g_{12}-jb_{12}=g_{13}-jb_{13}=1-j4$ and $g_{23}-jb_{23}=0.1-j0.4$.
}
\label{tab:solns3bus}
\end{table}

\captionsetup[figure]{font=small,skip=2pt}

Now we show that a partial Lagrangian behaves qualitatively differently. Suppose that we choose a strict local solution of \eqref{eqn:three_bus}. Let the multipliers corresponding to the equality constraints \eqref{eqn:3busb1} and \eqref{eqn:3busb2} be $\mu_1$ and $\mu_2$, respectively. The partial Lagrangian for \eqref{eqn:three_bus} is:
\begin{subequations} 
\begin{align} 
    &\mathcal{L}_{\mu}=  c(\sum_{j=2,3} g_{1j}-g_{1j}\cos(\theta_j)+b_{1j}\sin(\theta_j))\\
    &+  \mu_1( l_{2}+\sum_{j=1,3}( g_{2j}-g_{2j}\cos(\theta_{2j})-b_{2j}\sin(\theta_{2j})) ) \\
    &+  \mu_2( l_{3}+ \sum_{j=1,2}( g_{3j}-g_{3j}\cos(\theta_{3j})-b_{3j}\sin(\theta_{3j})) ).
\end{align}
\end{subequations}
In contrast to the penalized problem, there is only a single solution for $\mathcal{L}_{\mu}$ that satisfies both the first order and second order optimality conditions. It is close to the global solution of $\mathcal{L}_\rho$ (the black dot in Fig.~\ref{fig:3busHessian}(b)), even though the multipliers used in forming $\mathcal{L}_{\mu}$ are from a strict local solution.  

If we look at the Hessian of $\mathcal{L}_{\mu}$, we see that the Hessian is either negative definite or indefinite at the strict local solutions of $\mathcal{L}_{\rho}$ (the definiteness of the Hessians for $\mathcal{L}_{\mu}$ at the local solutions of $\mathcal{L}_{\rho}$ are listed in Table.~\ref{tab:solns3bus}). If the Hessian is not positive semidefinite, there is always a direction to lower the objective value of a function. For example, these descent directions are shown in Fig.~\ref{fig:3busHessian}(d) and Fig.~\ref{fig:3busHessian}(f). 

All together, Fig.~\ref{fig:3busHessian} shows how Algorithm 1 would get around the strict local solutions in $\mathcal{L}_{\rho}$.
Suppose we solve the Lagrangian from a point around the global minimum $\bx^{\star}$. Since $\nabla^{2} \mathcal{L}_{\mu}(\bx^{\star})$ is positive definite, this means the starting point is
at a valley of the Lagrangian surface. So solving the Lagrangian would return the global solution. Now let us we use a point around the local solution, say $\bar{\bx}$, as an initial point to solve the Lagrangian. 
As shown in (d) and (f) of Fig.~\ref{fig:3busHessian},
$\nabla^{2} \mathcal{L}_{\mu}(\bar{\bx})$ is negative definite or indefinite, so the surface of the Lagrangian is concave down or has a saddle. Then we can find at least one descent direction to get out of being trapped at the current point. 

\section{Analysis of Algorithm~1}
\label{sec:analysis}
In this section, we provide a more rigorous analysis of Algorithm~1 and give formal proofs. Particularly, we focus attention on systems with a tree topology. Readers who are interested in the simulation results for general meshed networks can proceed to the next section. 

We first consider a tree network with fixed voltage magnitudes and shown that the minimizer of the Lagrangian falls into the attraction basin of the global minimum of the ACOPF problem, which generalizes the observations in Section \ref{sec:geometry2bus}. 
Then we optimize over both voltage magnitudes and angles for a 2-bus network, and look at the Hessian matrix of the Lagrangian as we do in Section \ref{sec:three_bus}.  We proof that the Hessian matrix of the Lagrangian is positive definite at the global minimum and negative definite or indefinite at the local minimum. 

\subsection{Fixed voltage magnitudes}\label{sec:analysisA}
In this part, we consider a tree network with fixed voltage magnitudes. Particularly, we assume that the NLP solver runs a gradient descent-like algorithm. This means starting the solver from an initial point in the attraction basin of a solution would return this solution.
Formally, the attraction basin of a solution is defined as~\cite{bertsekas1997nonlinear}:
\begin{definition}\label{def:basin}
Let $\bx^{\star}$ be an unconstrained local minimum to $f:\R^{n}\longrightarrow\R$.
Assume there exists a set $\mathcal{X}$ such that $f(\bx)$ is continuously differentiable on $\mathcal{X}$ and $\bx^{\star}\in\mathcal{X}$.
For every point $\bx\neq\bx^{\star}$ and $\bx\in\mathcal{X}$, if the following inequality holds, then $\mathcal{X}$ is a subset of the attraction basin of $\bx^{\star}$:
\begin{align}\label{ineq:defbasin}
    \nabla f(\bx)^T(\bx^{\star}-\bx)< 0,~\forall\bx\neq\bx^{\star}, \bx\in\mathcal{X},
\end{align}
where $\nabla f(\bx)$ represents the gradient of $f(\cdot)$ at the point $\bx$.
\end{definition}
Intuitively, the inequality in \eqref{ineq:defbasin} implies that the direction where the function values decreases (descent direction) is aligned with the negative gradient. 
Now we give the following theorem about the performance of Algorithm 1 for a tree network with fixed voltage magnitudes:
\begin{theorem}\label{them1}
Consider an $N$-bus network with a tree topology and fix the voltage magnitudes. If Algorithm 1 is initialized from a local minimum, then it will escape from this local minimum. 
\end{theorem}

\begin{proof}
We prove Theorem~\ref{them1} by showing that the minimizer of the Lagrangian falls into the attraction basin of the global minimum of the ACOPF problem.
Since the NLP solver follows a gradient-like algorithm as we assumed at the beginning of this part, starting the solver from the minimizer of the Lagrangian would reach the global minimum.

We do the proof by induction starting with a 2-bus network. 
The ACOPF problem for the 2-bus network is given in \eqref{eqn:two_bus} and its Lagrangian is given in \eqref{eqn:Lagrangian2bus}. We first study the solutions to \eqref{eqn:two_bus} by looking at the equality constraint $h(\theta)=l+g-g \cos(\theta)-b\sin(\theta)=0$. Its gradient can be written as
\begin{equation*}
    h^{\prime}(\theta) = g\cos(\theta)(\tan(\theta)-b/g).
\end{equation*}
Suppose $\theta\in(-\pi/2, 3\pi/2)$, then $h^{\prime}$ is zero at $\theta=\tan^{-1}(b/g)$. We also have:
\begin{subequations}\label{eqs:h}
\begin{align}
   h^{\prime}(\theta)<&0, \forall~\theta\in(-\frac{\pi}{2},\tan^{-1}(b/g))\label{ineq1:gradh}\\
   h^{\prime}(\theta)>&0, \forall~\theta\in(\tan^{-1}(b/g), \tan^{-1}(b/g)+\pi).
\end{align}
\end{subequations}
This means $\theta=\tan^{-1}(b/g)$ is a minima of $h(\theta)$.
Since for a feasible problem, the solution to $h(\theta)=0$ must exist within $(-\pi/2, 3\pi/2)$, then by the intermediate value theorem, there are two solutions to \eqref{eqn:two_bus}, which satisfy the following inequalities:
\begin{equation}\label{ineq:solnsA}
    -{\pi}/{2}<\theta^{\star}< \tan^{-1}(b/g)<\bar{\theta}<{3\pi}/{2},
\end{equation}
where $\theta^{\star}$ is the global minimum and $\bar{\theta}$ is the local minimum (see Appendix \ref{appendix:acopf} for more details). 

Now we use \eqref{ineq:defbasin} to show that the interval $(-\pi/2, \tan^{-1}(b/g))$ is a subset of the attraction basin of $\theta^{\star}$. Suppose $\rho$ is sufficiently large, then 
$\theta^{\star}$ can be also treated as the global minimum of the unconstrained penalized problem in \eqref{eqn:pentwobus}. Therefore, this
is equivalent to showing
\begin{equation}\label{ineq:isbasin}
    \mathcal{L}_{\rho}^{\prime}(\theta)^T({\theta}^{\star}-\theta)<0, \forall~\theta\in (-\frac{\pi}{2}, \tan^{-1}(b/g)).
\end{equation}
As $\rho$ is sufficiently large, the sign of $\mathcal{L}_{\rho}^{\prime}(\theta)$ is dominated by the gradient of the second term in \eqref{eqn:pentwobus}:
\begin{subequations}
\begin{align}
        \mathcal{L}_{\rho}^{\prime}(\theta)\approx& \rho(l+g-g \cos(\theta)-b\sin(\theta))(g\sin (\theta)-b\cos(\theta))\nonumber\\
        =&\rho h(\theta)h^{\prime}(\theta).\nonumber
\end{align}
\end{subequations}
For any $\theta\in (-\pi/2, \tan^{-1}(b/g))$, we have  $h^{\prime}(\theta)<0$ from \eqref{ineq1:gradh}, which means the function $h(\theta)$ is decreasing on the interval $(-\pi/2, \tan^{-1}(b/g))$.
Also, the global minimum $\theta^{\star}$ must satisfy $h({\theta}^{\star})=0$.
Therefore we have
\begin{subequations}
\begin{align*}
   h(\theta)>&0, \forall~\theta\in(-{\pi}/{2},{\theta}^{\star})\\
   h(\theta)<&0, \forall~\theta\in({\theta}^{\star},\tan^{-1}(b/g)).
\end{align*}
\end{subequations}
Then the inequality in \eqref{ineq:isbasin}  follows from above. By Definition \ref{def:basin}, the interval $(-\pi/2, \tan^{-1}(b/g))$ is a subset of the attraction basin of $\theta^{\star}$. 

To obtain the minimizer of $\mathcal{L}_{\mu}$, we write out the optimality condition of \eqref{eqn:Lagrangian2bus} for the primal-dual optimal solution $(\hat{\theta}, \hat{\mu})$:
\begin{equation}\label{eq:optimality}
    (c^{\prime}+\hat{\mu}) g \sin(\hat{\theta})+(c^{\prime}-\hat{\mu}) b \cos(\hat{\theta})=0,
\end{equation}
where $c^{\prime}$ is a shorthand for $c^{\prime}(g-g\cos(\hat{\theta})+b\sin(\hat{\theta}))$ and is the gradient of the cost function.
Suppose $\hat{\theta}\in(-{\pi}/{2}, {3\pi}/{2})$, then $\hat{\theta}$ solves
\begin{align}\label{eq:minimizer}
  \tan^{-1}(\frac{\hat{\mu}-c^{\prime}}{\hat{\mu}+c^{\prime}}b/g)+k\pi,~k=0, 1,
\end{align}
where the smaller value is the minimum of $\mathcal{L}_{\mu}$ and the larger one is the maximum  (see Appendix \ref{appendix:Lag} for more details). Let $\hat{\theta}$ be the minimum, which satisfies $ -\pi/{2}<\hat{\theta}< \tan^{-1}(b/g)$.
Since the interval $(-\pi/2, \tan^{-1}(b/g))$ is a subset of the attraction basin of $\theta^{\star}$.
no matter what initial point we start Algorithm 1 from, solving the Lagrangian gives us a solution lying in the attraction basin of the global minimum, which enables Algorithm 1 to get out of a strict local solution.

\captionsetup[figure]{font=small,skip=2pt}
\begin{figure}[t]
     \centering
    \begin{subfigure}[b]{0.5\columnwidth}
         \centering
         \includegraphics[scale=0.25]{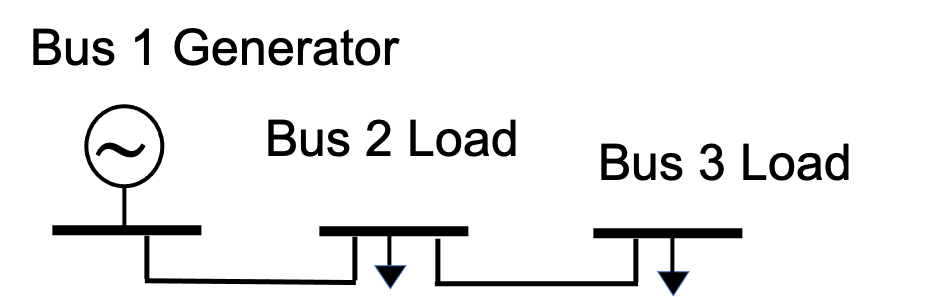}
         \caption{}
         \label{fig:typeA}
     \end{subfigure}
     \begin{subfigure}[b]{0.4\columnwidth}
         \centering
         \includegraphics[scale=0.25]{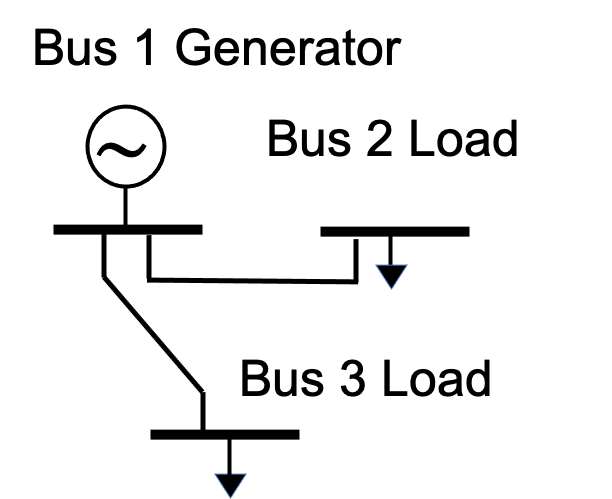}
         \caption{}
         \label{fig:typeB}
     \end{subfigure}
\caption{The two types of three bus networks with the tree structure.
        \vspace{-0.5cm}}
        \label{fig:3bustree}
\end{figure}

Now we induct from 2-bus to 3-bus networks. There are two types of tree topology for a 3-bus network, which are shown in Fig.~\ref{fig:3bustree}. Since the topology in Fig.~\ref{fig:3bustree}(b) is equivalent to two 2-bus networks, we focus on the 3-bus branch in Fig.~\ref{fig:3bustree}(a), where bus 1 is reference bus.

Since the cost function $c(\cdot)$ is increasing, given the load at bus 2 and bus 3, minimizing the power generation cost in \eqref{prob1} is equivalent to minimizing the power transfer cost on both lines. Suppose the load at bus 2 and bus 3 are $l_2$ and $l_3$, respectively, then the optimization problem is:
\begin{subequations} \label{eqn:3bus(a)}
\begin{align}
    \min_{\theta_{12},\theta_{23}} \; & c_1(P_{12})+
    c_2(P_{23})\\
    \st~& l_2+P_{21}+P_{23}=0\label{eq:(a)bus2}\\
    & l_3+P_{32}=0.\label{eq:(a)bus3}
\end{align}
\end{subequations}

With $l_2$ fixed,
for every given $\theta_{23}$, we can always pick some $\theta_{12}$ to satisfy \eqref{eq:(a)bus2}. Therefore, we can look at $\theta_{23}$ as the optimization variable first. Then the problem \eqref{eqn:3bus(a)} is reduced to
\begin{subequations}
\begin{align}
    \min_{\theta_{23}} \; &
    \tilde{c}(P_{23})\nonumber\\
    \st~& l_3+P_{32}=0\nonumber
\end{align}
\end{subequations}
where $\tilde{c}(\cdot)$ is some increasing cost function that takes into account the effect of $\theta_{23}$ on $P_{12}$. This problem has exactly the same formulation as the 2-bus network in \eqref{eqn:two_bus}. As we proved for the 2-bus network, if we start Algorithm 1 from a point where $\theta_{23}$ is at the local minimum, then we can get out of this local minimum.

Now we optimize $\theta_{12}$ for a given $\theta_{23}$, then we can add $P_{23}$ to $l_2$. Therefore, \eqref{eqn:3bus(a)} is reduced to
\begin{subequations}
\begin{align}
    \min_{\theta_{12}} \; & c_1(P_{12})\nonumber\\
    \st~& \tilde{l}_2+P_{21}=0\nonumber
\end{align}
\end{subequations}
where $\tilde{l}_2 = l_2+P_{23}$.
This problem also has the same formulation as the 2-bus network in \eqref{eqn:two_bus}. Therefore, if the initial point is a local minimum for $\theta_{12}$, Algorithm 1 can also get out of being stuck at this local minimum.

Let us assume Theorem \ref{them1} holds for a $(N-1)$-bus network and consider a $N$-bus network.
Similar to the proof for a 3-bus network, we can reduce the ACOPF problem for a $N$-bus network to the case of $(N-1)$ bus.
So by induction, Theorem \ref{them1} holds for the $N$-bus network.
\end{proof}

\subsection{Optimizing both voltage magnitudes and angles}\label{sec:analysisB}
In this part, we optimize both voltage magnitudes and angles for a 2-bus network. For simplicity, we ignore the reactive power. Suppose bus 1 is a generator and the reference (slack) bus with linear cost \$1/MW, and bus 2 is the load bus with load $l$. The ACOPF in \eqref{prob1} can be simplified as
\begin{subequations}\label{eqn:2bus_mag}
\begin{align}
\min_{\theta, V_{1}, V_{2}} & g V_{1}^{2}-V_{1} V_{2}(g \cos(\theta)-b \sin(\theta))\\
s.t. & l + V_{2}^{2} g-V_{1} V_{2}(g \cos(\theta)+b\sin (\theta))=0\label{eq:two_bus_mag_eq}\\
& V_{\min } \leq V_{1}, V_{2} \leq V_{\max }\label{eq:two_bus_mag_ineq}.
\end{align}
\end{subequations}
Let us collect all the variables into the vector $\bx=(\theta,V_1,V_2)$. We denote the objective function by $f(\bx)$, and the equality constraint \eqref{eq:two_bus_mag_eq} by $h(\bx)=0$.
The following theorem looks at the Hessian matrix of the Lagrangian.

\begin{theorem}\label{them:2busmag}
Denote the global solution of \eqref{eqn:2bus_mag} as $\bx^{\star}$ and the local solution as $\bar{\bx}$.
Then the Hessian matrix of a Lagrangian of \eqref{eqn:2bus_mag}, formed with multipliers at any of the local solutions, is positive definite at $\bx^{\star}$ and negative definite or indefinite at $\bar{\bx}$.
\end{theorem}

\begin{proof}
To study the solution to \eqref{eqn:2bus_mag}, we look at the equality constraint \eqref{eq:two_bus_mag_eq} directly. Its gradient with respect to $\theta$ can be written as
\begin{align*}
{\partial h}/{\partial \theta}=g \sin(\theta)-b \cos(\theta)=g \cos \theta(\tan(\theta)-b/g).
\end{align*}
Suppose $\theta\in(-\pi/2,3\pi/2)$, then ${\partial h}/{\partial \theta}$ is zero at $\tan ^{-1}(b/g)+k \pi$, $k=0,~1$, where the smaller value is located at the global minimum and the larger value is at the local minimum. Denote the global minimum as $\theta^{\star}$ and at the local minimum as $\bar{\theta}$. They satisfy (see Appendix \ref{appendix:acopf} for the details):
\begin{align}\label{ineq:solns}
    -{\pi}/{2}<\theta^{\star}<\tan ^{-1}(b/g)<\bar{\theta}<{3\pi}/{2}.
\end{align}

In Appendix~\ref{appendix:inactivity}, we show that at least one of $V_1$ and $V_2$ need to be binding at a constraint, but both voltages cannot be binding at the same time. This allows us to consider the cases where $V_1$ is binding or $V_2$ is binding separately.

{First, suppose $V_1$ is inactive and $V_2$ is binding.} In this case, $V_2$ is a constant and the Lagrangian of \eqref{eqn:2bus_mag} can be written as
\begin{align*}
\mathcal{L}_{\lambda, \bmu} =& f(\bx)+\mu h(\bx)+ \bar{\lambda}_{1}\left(V_{1}-V_{\max }\right)+\underline{\lambda}_{1}\left(-V_{1}+V_{\min }\right). 
\end{align*}
The multipliers are associated with some local solution, and $\mu$ is the Lagrange multiplier related to the equality constraint, and $\bar{\lambda}_{1}$ and $\underline{\lambda}_{1}$ are the multipliers related to the inequality constraints of $V_1$. 

Denote the Hessian matrix of $\mathcal{L}_{\lambda, \bmu}$ as $\nabla^{2}\mathcal{L}_{\lambda, \bmu}(\bx)$. To determine its definiteness, we write out all the leading principal minors at a solution $\tilde{\bx}$ (see Appendix \ref{appendix:Hessian} for the details):
\begin{subequations}\label{eq:case1}
\begin{align}
    D_{1}(\tilde{\bx})&=\tilde{V}_1\tilde{V}_2\frac{-2 g b}{g \cos(\tilde{\theta})(\tan (\tilde{\theta})-b/g)}\\
    D_{2}(\tilde{\bx})&= 2gD_{1}(\tilde{\bx}).
\end{align}
\end{subequations}
Following from the inequalities in \eqref{ineq:solns}, both leading principal minors in \eqref{eq:case1} are positive at the global minimum and negative at the local minimum. This means the Hessian matrix at ${\bx}^{\star}$ is positive definite.
In contrast,  the Hessian matrix at $\bar{\bx}$ is negative definite.

Now we suppose $V_2$ is inactive and $V_1$ is binding. In this case, $V_1$ is a constant and the Lagrangian is:
\begin{align*}
\mathcal{L}_{\lambda, \bmu} =& f(\bx)+\mu h(\bx)+ \bar{\lambda}_{2}\left(V_{2}-V_{\max }\right)+\underline{\lambda}_{2}\left(-V_{2}+V_{\min }\right). 
\end{align*}
Where the multipliers are associated with some local solution. 
Let us denote the Hessian matrix of the Lagrangian as $\tilde{\nabla}^{2}\mathcal{L}_{\lambda, \bmu}(\bx)$. Its leading principal minors at a feasible solution $\tilde{\bx}$ are (see Appendix \ref{appendix:Hessian} for the details):
\begin{subequations}\label{eq:case2}
\begin{align}
    D_{1}(\tilde{\bx})&=\tilde{V}_1\tilde{V}_2\frac{-2 g b}{g \cos(\tilde{\theta})(\tan (\tilde{\theta})-b/g)}\\
    \tilde{D}_{2}(\tilde{\bx})&= 2g\tilde{\mu} D_{1}(\tilde{\bx}).
\end{align}
\end{subequations}
Since the multiplier $\mu$ represents the marginal price of consuming each additional unit of load, it is positive at the global minimum. This means 
$D_{2}(\tilde{\bx})$ has the same sign as $D_{1}(\tilde{\bx})$. For the global minimum ${\bx}^{\star}$, $D_{1}(\tilde{\bx})$ is positive from \eqref{ineq:solns}, hence both leading principal minors in \eqref{eq:case2} are positive and the Hessian matrix is positive definite at ${\bx}^{\star}$. In contrast, at the local minimum $\bar{\bx}$, $D_{1}(\tilde{\bx})$ is negative following from \eqref{ineq:solns}. Then the Hessian matrix is either negative definite or indefinite at $\bar{\bx}$.
\end{proof}

The simulation results in the next section do not need to make any of the assumptions in Theorem~\ref{them1} and~\ref{them:2busmag}. They are about mesh networks with all constraints included. Therefore, we suspect the theory can be made much stronger and would extend to larger meshed networks. However, analyzing these cases is challenging and is a future direction for us. 

\vspace{-0.2cm}
\section{Simulation Results} \label{sec:results}
In this section we report the simulation results to validate the effectiveness of our algorithm.
The NLP solver used here is IPOPT \cite{wachter2006implementation} and the convergence tolerance is set to $0.0001$. It returns a feasible solution, which may or may not be a global optimum.
We test our algorithm on IEEE networks with 3, 9, 22, and 39 buses.
For the 3-bus, 9-bus and 22-bus networks, the local and global solutions are known and listed in~\cite{bukhsh2013local,nguyen2014appearance}. We use the strict local solutions as starting points for the solver to demonstrate the ability of Algorithm 1 of getting out of local solutions. For the 39-bus network, we do an exhaustive search by discretizing each variable within their bounds to find the global solution.
The simulation results show that for the 3, 9 and 22-bus networks, Algorithm 1 finds the globally optimal solution in $1$ iteration. For the 39-bus networks, it takes at most $3$ iterations for Algorithm 1 to obtain the optimal solution.

\subsection{3-Bus Network}
The three bus network we use is shown in Fig.~\ref{fig:typeA} and the voltage bounds are $[0.95, 1.05]$.
Two solutions exist and they are listed in Table~\ref{tab:solns3busA}. This was an example used in~\cite{nguyen2014appearance} to show that multiple reasonably looking local solutions can exist, and contrary to conventional wisdom, the higher voltage one is the suboptimal one (although the cost differences is small).

If we start the nonlinear solver from an initial point near the second solution, then the solver cannot get out of the attraction basin and always returns the second solution. In contrast, if we launch Algorithm~1 using the second solution as a starting point, then the algorithm converges to the first solution (the global solution) after one iteration. Although the cost difference is small between the two solutions, larger networks will have bigger cost differences. 

\begin{table}[ht]
\captionsetup{font=small}
\normalsize
\begin{tabular}{lllll}
\hline
 & Bus 1 & Bus 2 & Bus 3 & Cost \\
\hline
Solution 1 & $0.95\angle{0}$ & $0.95\angle{-0.48}$ & $0.98\angle{-0.53}$ & $1$\\
Solution 2 & $0.95\angle{0}$ & $1.01\angle{-0.46}$ & $1.05\angle{-0.51}$ & $1.0021$\\
\hline
\end{tabular}
\caption{The two local solutions for the three bus network in Fig.~\ref{fig:typeA}. The cost is normalized to 1 for the global solution.}
\label{tab:solns3busA}
\end{table}

\subsection{9-Bus Network}
\begin{figure}[t]
\centering
\includegraphics[height=4cm, width=5cm]{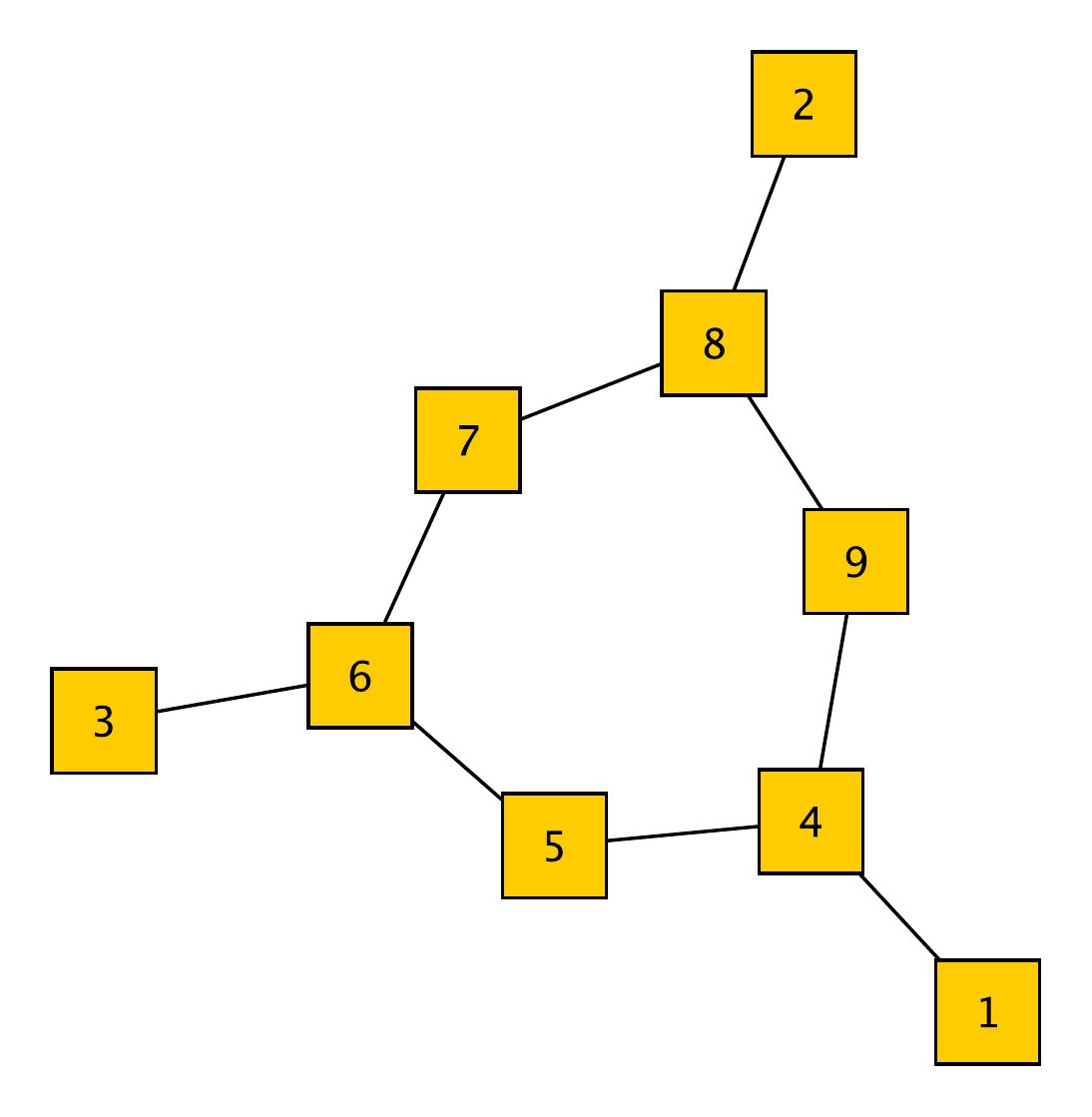}
\caption{Topology diagram of the nine-bus network.
\vspace{-0.5cm}
}
\label{fig:9bus}
\end{figure}

The topology of the 9-bus network is shown in Fig.~\ref{fig:9bus}. There are $3$ generators (bus 1, 2 and 3) and $9$ transmission lines. The voltage bounds are $[0.9, 1.1]$. Four solutions exist. The cost of the worst local solution is $38\%$ more than the cost at the global solution. 
We also find that the solutions at generators 2 and 3 and load buses 6, 7, and 8 are important to improve the cost.
The power transfer along the lines between these buses tend to get stuck at a suboptimal solution, which leads to a cost more than $30\%$ higher than the lowest one. For the nonlinear solver, we need to relaunch it using different initial points in order for these five nodes to get around the attraction basin.
This requires many trials. 
In contrast, 
Algorithm~1 only requires one iteration to achieve the global solution, even starting from the local solution with the highest cost. 
\captionsetup[figure]{font=small,skip=2pt}

\begin{table*}[t]
\captionsetup{font=small}
\normalsize
\centering
\begin{tabular}{lllllll}
\hline
& Bus 2 & Bus 7 & Bus 12 & Bus 17 & Bus 22 & cost\\
\hline
Solution 1 & $1.0285\angle{-0.045}$& $1.05\angle{0}$& $1.0285\angle{-0.045}$& $1.05\angle{0}$&$1.0285\angle{-0.045}$& $1$\\
Solution 2 & $0.95\angle{-0.339}$& $1.0145\angle{4.57}$& $0.95\angle{3.089}$& $1.0145\angle{1.714}$& $0.95\angle{0.233}$& $1.306$\\
\hline
\end{tabular}
\caption{The two solutions for the 22-bus network. We pick five buses and show their voltage and angles. The costs at the two solutions are normalized such that the globally optimal cost is $1$.}
\label{tab:solns22bus}
\end{table*}

\subsection{22-Bus Network}
In the 22-bus network, the buses are connected in a loop. There are 11 generators and 22 transmission lines. The voltage bounds are $[0.95,1.05]$. There exist two solutions, and the cost of the local solution is $30\%$ higher than that of the global solution. The two solutions are quite different. We pick $5$ buses that are evenly spaced and list their solutions in Table \ref{tab:solns22bus}. Since the two solutions are very different, it is hard for a nonlinear solver to get around the local solution. 

Particularly, if we initialize the solver with a flat start, we obtain the strict local solution. Furthermore, we generate $100$ random points uniformly at random within the bounds of each variable. If these points are used to initiate the nonlinear solver, the local solution is always obtained and the global one cannot be reached. In comparison, Algorithm~1 can achieve the global solution after one iteration regardless of the initial point. This is an example where using random search is very computationally inefficient, and our deterministic algorithm turns out to be much more successful.

\subsection{39-Bus Network}
In the 39 bus network, there are 10 generators and 46 transmission lines. The voltage bounds are $[0.95, 1.05]$. 
Unlike the previous smaller networks, the number and the cost of the solutions are not previously known for this network. Therefore we conducted an exhaustive search to find the global solution. 
To evaluate the effectiveness of Algorithm~1, we choose $600$ random points within the bounds of each variable using the uniform distribution. Then we start Algorithm~1 with these random points to observe the improvement of the solution quality. 

\begin{figure}[t]
\centering
\includegraphics[height=5cm, width=6cm]{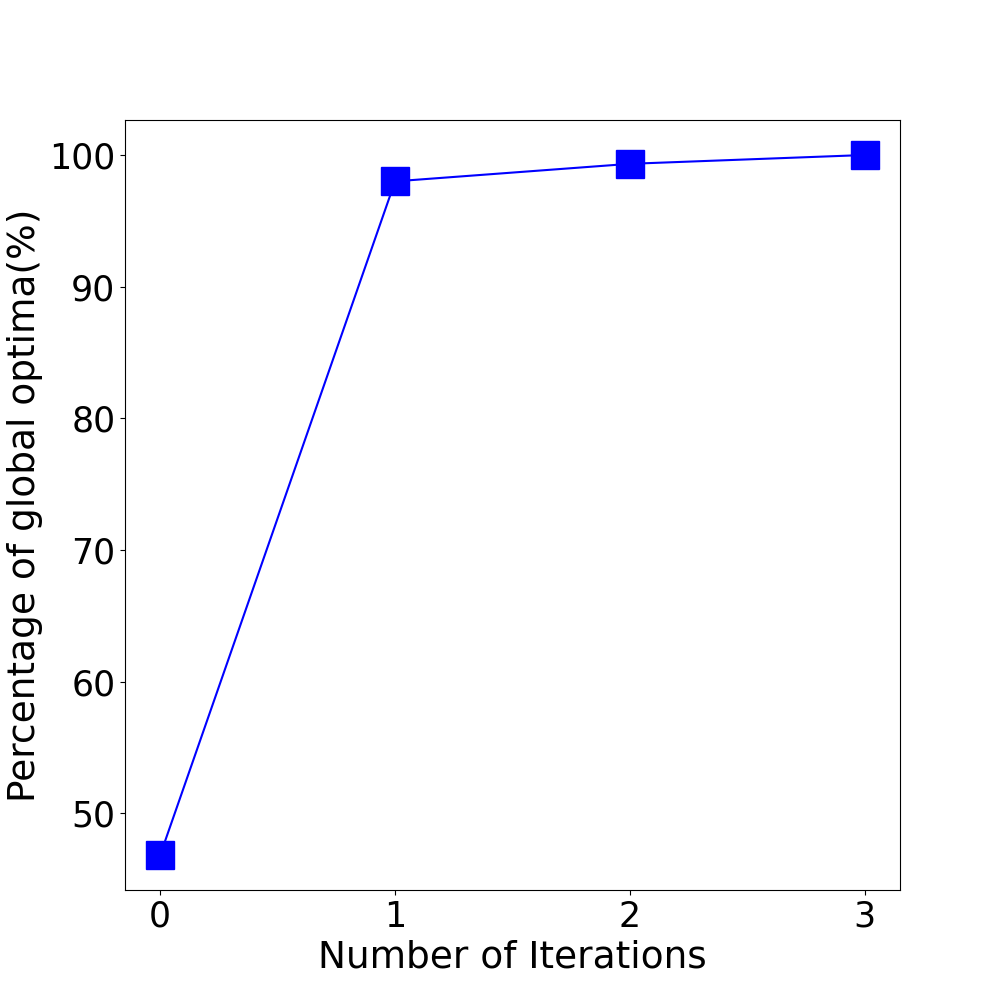}
\caption{Using a set of random starting points, 47\% of them leads to the global optimal after a direct call to IPOPT. The fraction of global optimal solutions increases to 98\%, 99.93\% and 100\% after one, two and three iterations of algorithm 1, respectively. 
\vspace{-0.5cm}}
\label{fig:39busfrac}
\end{figure}

In Fig. \ref{fig:39busfrac}, we plot the fraction of global solutions in the set of all $600$ results after each iteration.
The x-axis represents the number of iterations that Algorithm 1 is ran, and y-axis represents the percentage of globally optimal solutions after each iteration. When we make a direct call to the solver, 
less than half of the solutions are globally optimal. 
One application of Algorithm 1 increases the percentage of globally optimal solutions to $98\%$.
After two iterations, only four cases are not globally optimal. When we run Algorithm~1 for three iterations, all
solutions are globally optimal.

\begin{figure}[t]
\centering
\includegraphics[height=5cm, width=6cm]{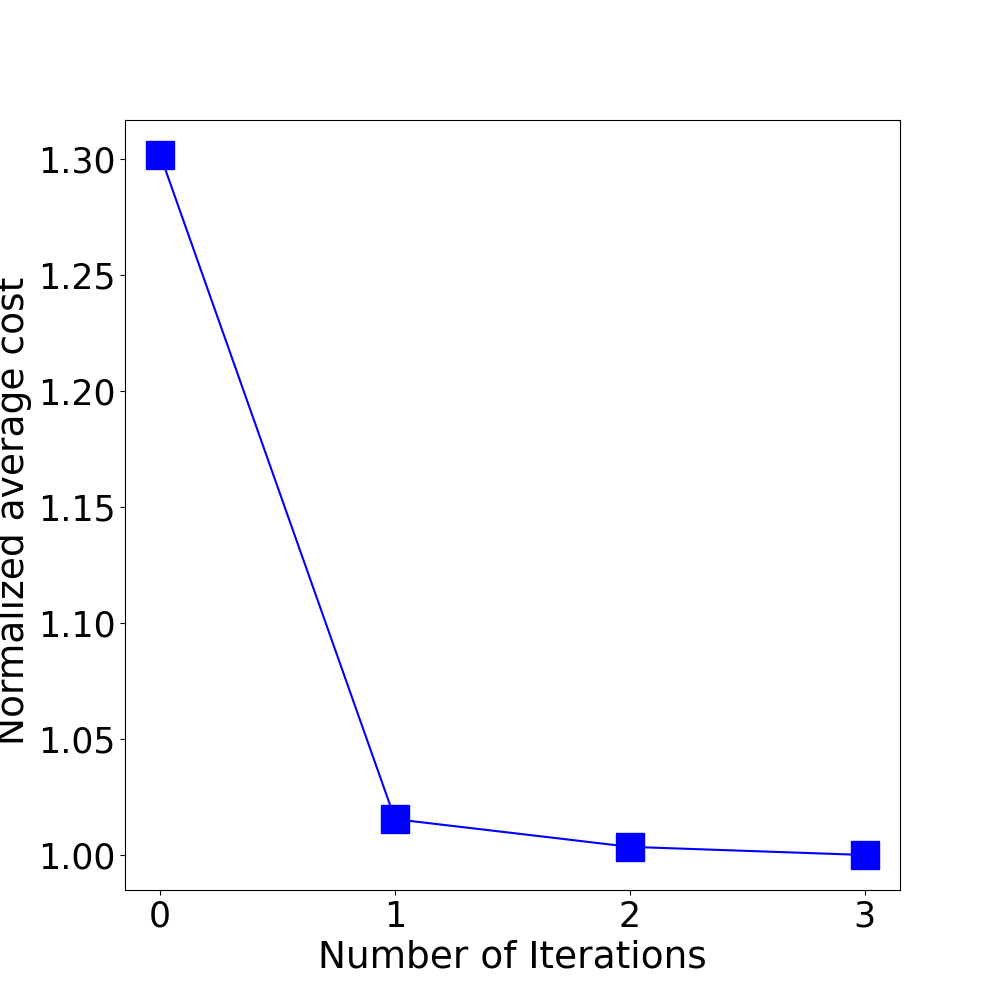}
\caption{The average cost of all $600$ solutions, and is normalized such that the optimal cost is $1$. After a direct call to IPOPT, the average cost is $30\%$ higher than the optimal cost. Then the average cost
reduces to $1.5\%$, $0.4\%$ higher than the optimal value after one and two iterations of Algorithm 1, respectively. After three iterations, the average cost is exactly the optimal cost.
\vspace{-0.5cm}}
\label{fig:39buscost}
\end{figure}

We also calculate the average cost of the $600$ solutions after each iteration of Algorithm~1 and plot the result in Fig.~\ref{fig:39buscost}. The x-axis is the number of iterations of running Algorithm 1, and y-axis represents the average cost of $600$ solutions, which is normalized using the optimal cost as the factor. After a direct call to the solver, the average cost is $30\%$ higher than the optimal cost. As Algorithm~1 is ran, the average cost decreases quickly. After one iteration, the average cost is only $1.5\%$ more than the globally optimal cost, and after three iterations all solution are at the global optimum.

\captionsetup[figure]{font=small,skip=2pt}

\captionsetup[figure]{font=small,skip=2pt}

\section{Conclusion}
\label{sec:conclusion}
In this paper, we propose a simple algorithm to iteratively improve the solution quality of ACOPF problems.
First, we solve the ACOPF problem using an existing nonlinear solver. From the solution and its associated dual variables, we construct a partial Lagrangian. Optimizing this partial Lagrangian leads to a new solution. With this solution as an initial point, we again call the solver for the ACOPF problem. By repeating these steps, we can iteratively improve the solution quality, escaping from local solutions to find better ones. We illustrate the intuition behind our algorithm using 2 and 3-bus networks, which shows that the partial Lagrangian has a flatter optimization landscape compared to the original primal problem. We prove the algorithm is guaranteed to work in tree networks. 
We validate the effectiveness of our algorithm on standard 9-bus, 22-bus and 39-bus networks. 
Regardless of the initial points, our algorithm always finds the global optimum within at most three iterations.

\bibliographystyle{IEEEtran}
\bibliography{mybib.bib}

\appendix
\subsection{Determine global minimum for ACOPF}\label{appendix:acopf}
In Section \ref{sec:analysisA} and \ref{sec:analysisB}, we find two solutions to the supply/balance equality constraint, which satisfy the inequalities in \eqref{ineq:solnsA} or \eqref{ineq:solns}. 
In this part, we give the reason why the smaller solution in \eqref{ineq:solnsA} (or \eqref{ineq:solns}) is the global minimum and the larger solution is the local minimum.

Let us subtract the power received at the load bus from the generation at the the generator, then we have the transmission loss as follows:
\begin{align}
    \text{loss}&=g-g\cos(\theta)+b\sin(\theta)-(-g+g\cos(\theta)+b\sin(\theta))\nonumber\\
    &=2g(1-\cos(\theta)).\nonumber
\end{align}
Due to the periodicity of arctangent function, the larger value $\bar{\theta}$ must be larger than $\pi/2$. Then the loss at $\theta^{\star}$ is smaller than the loss at $\bar{\theta}$. So $\theta^{\star}$ is an more optimal solution than $\bar{\theta}$. Since there are only two solutions for this example, $\theta^{\star}$ must be the global minimum and $\bar{\theta}$ is the strict local minimum.

\subsection{Determine global minimum for the Lagrangian}\label{appendix:Lag}
In this part, we determine the global minimum of the Lagrangian problem for the 2-bus network, where we fix the voltage magnitudes and optimize the angles. 

Let us denote the two solutions of the Lagrangian problem in \eqref{eqn:Lagrangian2bus} as $\hat{\theta}$ and $\bar{\theta}$, and the multipliers associated with them are $\hat{\mu}$ and $\bar{\mu}$, respectively. Then from \eqref{eq:minimizer}, we have
\begin{subequations}\label{ineq:minimizer}
\begin{align}
    &-\frac{\pi}{2}<\hat{\theta}=\tan^{-1}(\frac{\hat{\mu}-c^{\prime}}{\hat{\mu}+c^{\prime}}\frac{b}{g})<\tan^{-1}(\frac{b}{g})\\
     &\frac{\pi}{2}<\bar{\theta}=\tan^{-1}(\frac{\bar{\mu}-c^{\prime}}{\bar{\mu}+c^{\prime}}\frac{b}{g})+\pi<\tan^{-1}(\frac{b}{g})+\pi.
\end{align}
\end{subequations}
Also we can represent the multiplier using $\theta$ by rearranging the terms in \eqref{eq:optimality}.
We take $(\hat{\theta},\hat{\mu})$ as an example, and $\bar{\mu}$ can be represented using $\bar{\theta}$ in a similar way. The expression of $\hat{\mu}$ in terms of $\hat{\theta}$ is
\begin{align}\label{eq:expressmu}
    \hat{\mu}=-\frac{g \sin(\hat{\theta})+b \cos(\hat{\theta})}{g \sin(\hat{\theta})-b \cos(\hat{\theta})}.
\end{align}
Now let us write out the second-order derivative of the Lagrangian function, and plug \eqref{eq:expressmu} into it. Then we have:
\begin{align}
    \mathcal{L}_{\hat{\mu}}^{\prime \prime}(\hat{\theta})&=(1+\hat{\mu}) g \cos (\hat{\theta})-(1-\hat{\mu}) b \sin(\hat{\theta})\nonumber\\
    &=-\frac{2gb}{g \cos(\hat{\theta})(\tan (\hat{\theta})-\frac{b}{g})}.\nonumber
\end{align}
Using the inequalities in \eqref{ineq:minimizer}, we have
\begin{align}
     &\mathcal{L}_{\hat{\mu}}^{\prime \prime}(\hat{\theta})>0,\nonumber\\
      &\mathcal{L}_{\bar{\mu}}^{\prime \prime}(\bar{\theta})<0.\nonumber
\end{align}
This means that $\hat{\theta}$ is the minimum of the Lagrangian problem, and $\bar{\theta}$ is the maximum.

\subsection{Inactivity of inequality constraints in \eqref{eq:two_bus_mag_ineq}}
\label{appendix:inactivity}
In this part, we prove that not all inequality constraints in \eqref{eq:two_bus_mag_ineq} are inactive by contradiction.
We first suppose all inequality constraints in \eqref{eq:two_bus_mag_ineq} are inactive, and convert \eqref{eqn:2bus_mag} to the penalized unconstrained formulation: 
\begin{equation}\label{eqn:2bus_mag_pen}
\mathcal{L}_{\rho}(\bx) = f(\bx)+\rho/2 [h(\bx)]^{2}.
\end{equation}
Assume $\rho$ is sufficiently large, then \eqref{eqn:2bus_mag_pen} can be viewed as being equivalent to the original problem \eqref{eqn:2bus_mag}. Let us take gradients of $\mathcal{L}_{\rho}(\bx)$ with respect to $V_1$ and $V_2$ at a feasible solution $\tilde{\bx}$. 
Since $\tilde{\bx}$ satisfies $h(\tilde{\bx})=0$,
the terms multiplied by $\rho h$ in the gradients can be ignored. So the gradients are given by
\begin{subequations}
\begin{align}
\frac{\partial\mathcal{L}_{\rho}}{\partial V_{1}}&=2g \tilde{V}_{1}-\tilde{V}_{2}(g \cos(\tilde{\theta})-b \sin(\tilde{\theta}))\label{eq:grad_V1}\\
\frac{\partial\mathcal{L}_{\rho}}{\partial V_{2}}&=-\tilde{V}_{1}(g \cos(\tilde{\theta})-b \sin(\tilde{\theta})).\label{eq:grad_V2}
\end{align}
\end{subequations}\\
1) If $\frac{\partial\mathcal{L}_{\rho}}{\partial V_{1}}=0$, then we have
\begin{align}
g \cos(\tilde{\theta})-b \sin(\tilde{\theta})=2 g \frac{\tilde{V}_{1}}{\tilde{V}_{2}}~(\tilde{V}_{2}\neq 0).\label{eq:fmgrad_V1}
\end{align}
Plug \eqref{eq:fmgrad_V1} into \eqref{eq:grad_V2} and we get
\begin{align}
\frac{\partial\mathcal{L}_{\rho}}{\partial V_{2}}=-2g \frac{\tilde{V}_{1}^2}{\tilde{V}_{2}}<0.\nonumber
\end{align}
This means if $V_1$ is inactive, then $V_2$ must be on the boundary of the constraint set.\\
2) Suppose $\frac{\partial\mathcal{L}_{\rho}}{\partial V_{2}}=0$. Since $\tilde{V}_1\neq 0$, we have
\begin{align}
g \cos(\tilde{\theta})-b \sin(\tilde{\theta})=0.\label{eq:fmgrad_v1}
\end{align}
If we plug \eqref{eq:fmgrad_v1} into \eqref{eq:grad_V1}, then we have
\begin{align}
\frac{\partial\mathcal{L}_{\rho}}{\partial V_{1}}=2g \tilde{V}_{1}>0.\nonumber
\end{align}
That is, if $V_2$ is inactive, then $V_1$ must be on the boundary of the constraint set. Therefore one of $V_1$ and $V_2$ must be binding, and
\eqref{eqn:2bus_mag} can be reduced to the bivariate optimization problem. 

\subsection{Hessian matrix of the Lagrangian}\label{appendix:Hessian}
In this part, we derive the Hessian matrix of the Lagrangian function for problem \eqref{eqn:2bus_mag}, where we optimize both voltage magnitudes and angles for a 2-bus network.
In Appendix \ref{appendix:inactivity}, we have shown that one of $V_1$ and $V_2$ must be binding, so here we consider the cases where $V_1$ is binding or $V_2$ is binding separately.

We first suppose $V_1$ is inactive ($V_2$ is binding).
Then the Hessian matrix of the Lagrangian is
\begin{align}
    \nabla^{2}\mathcal{L}_{\lambda, \bmu}=\left(\begin{array}{cc}
\frac{\partial^{2} \mathcal{L}_{\lambda, \bmu}}{\partial \theta^{2}} & \frac{\partial^{2} \mathcal{L}_{\lambda, \bmu}}{\partial \theta \partial V_{1}} \\
\frac{\partial^{2} \mathcal{L}_{\lambda, \bmu}}{\partial \theta \partial V_{1}} & \frac{\partial^{2} \mathcal{L}_{\lambda, \bmu}}{\partial V_{1}^{2}}
\end{array}\right).\nonumber
\end{align}
The two leading principal minors of $\nabla^{2}\mathcal{L}_{\lambda, \bmu}$ at a feasible solution $\tilde{\bx}$ are
\begin{subequations}
\begin{align}
    D_1(\tilde{\bx}) & = \frac{\partial^{2} \mathcal{L}_{\lambda, \bmu}}{\partial \theta^{2}}\nonumber\\
    &=\tilde{V}_{1} \tilde{V}_{2}[(1+\tilde{\mu}) g \cos(\tilde{\theta})-(1-\tilde{\mu}) b \sin(\tilde{\theta})]\nonumber\\
    D_2(\tilde{\bx}) & = \nabla^{2}\mathcal{L}_{\lambda, \bmu}\nonumber\\
    &=2gD_1(\tilde{\bx})-\tilde{V}_{2} ^{2}[(1+\tilde{\mu}) g \sin(\tilde{\theta})+(1-\tilde{\mu}) b \cos(\tilde{\theta})]^2\nonumber
\end{align}
\end{subequations}
where $\tilde{\mu}$ is the dual solution associated with $\tilde{\bx}$. If $(\tilde{\bx}, \tilde{\mu})$ are also the optimal solution, then we can write out the 
optimality condition of the Lagrangian for $\theta$:
\begin{equation}\label{eq:optimality_theta}
    (1+\tilde{\mu}) g \sin(\tilde{\theta})+(1-\tilde{\mu}) b \cos(\tilde{\theta})=0.
\end{equation}
From \eqref{eq:optimality_theta}, $D_2(\tilde{\bx})$ can be simplified as 
\begin{align}
D_{2}(\tilde{\bx})&=2gD_1(\tilde{\bx})\nonumber.
\end{align}
Also, we can represent $\tilde{\mu}$ in terms of $\tilde{\theta}$:
\begin{align}\label{eq:expressmuC}
   \tilde{\mu}=-\frac{g \sin(\tilde{\theta})+b \cos(\tilde{\theta})}{g \sin(\tilde{\theta})-b \cos(\tilde{\theta})}.
\end{align}
If we plug \eqref{eq:expressmuC} into $D_{1}(\tilde{\bx})$, then we have
\begin{align}
D_{1}(\tilde{\bx})&=\tilde{V}_1\tilde{V}_2\frac{-2 g b}{g \cos(\tilde{\theta})(\tan(\tilde{\theta})-\frac{b}{g})}\nonumber.
\end{align}
Following from the inequalities in \eqref{ineq:solns}, both $D_1(\tilde{\bx})$ and $D_2(\tilde{\bx})$ are positive at the global minimum and negative at the local minimum. Hence Theorem \ref{them:2busmag} holds for the case where $V_1$ is inactive and $V_2$ is binding.

Now we suppose $V_2$ is inactive ($V_1$ is binding).
Then the Hessian matrix of the Lagrangian  is
\begin{align}
   \tilde{\nabla}^{2}\mathcal{L}_{\lambda, \bmu}=\left(\begin{array}{cc}
\frac{\partial^{2} \mathcal{L}_{\lambda, \bmu}}{\partial \theta^{2}} & \frac{\partial^{2} \mathcal{L}_{\lambda, \bmu}}{\partial \theta \partial V_{2}} \\
\frac{\partial^{2} \mathcal{L}_{\lambda, \bmu}}{\partial \theta \partial V_{2}} & \frac{\partial^{2} \mathcal{L}_{\lambda, \bmu}}{\partial V_{2}^{2}}
\end{array}\right).
\end{align}
The two leading principal minors of $\tilde{\nabla}^{2}\mathcal{L}_{\lambda, \bmu}$ at a feasible solution $\tilde{\bx}$ are
\begin{subequations}
\begin{align}
    D_1(\tilde{\bx}) & = \frac{\partial^{2} \mathcal{L}_{\lambda, \bmu}}{\partial \theta^{2}}\nonumber\\
    &=\tilde{V}_1\tilde{V}_2\frac{-2 g b}{g \cos (\tilde{\theta})(\tan(\tilde{\theta})-\frac{b}{g})}\\
    \tilde{D}_2(\tilde{\bx}) & = \tilde{\nabla}^{2}\mathcal{L}_{\lambda, \bmu}\nonumber\\
    &=2g \tilde{\mu}D_1(\tilde{\bx})-\tilde{V}_{1} ^{2}[(1+\tilde{\mu}) g \sin(\tilde{\theta})+(1-\tilde{\mu}) b \cos (\tilde{\theta})]^2
\end{align}
\end{subequations}
where $\tilde{\mu}$ is the dual solution associated with $\tilde{\bx}$.
If $(\tilde{\bx}, \tilde{\mu})$ are also the optimal solutions, then the optimality condition in \eqref{eq:optimality_theta} still holds and $\tilde{D}_2(\tilde{\bx})$ can be simplified as
\begin{subequations}
\begin{align}
    \tilde{D}_2(\tilde{\bx})=2g \tilde{\mu}D_1(\tilde{\bx}).
\end{align}
\end{subequations}
Since the multiplier $\tilde{\mu}$ represents the marginal price and is positive at the global minimum, $\tilde{D}_2(\tilde{\bx})$ has the same sign as $\tilde{D}_1(\tilde{\bx})$.
From the inequalities in \eqref{ineq:solns},
$\tilde{D}_1(\tilde{\bx})$ is positive, hence the Hessian matrix $\tilde{\nabla}^{2}\mathcal{L}_{\lambda, \bmu}$ is positive definite at the global minimum. For the local minimum, since $\tilde{D}_1(\tilde{\bx})$ is negative from \eqref{ineq:solns}, the Hessian matrix $\tilde{\nabla}^{2}\mathcal{L}_{\lambda, \bmu}$ cannot be positive definite. This means it is either negative definite or indefinite at the local minimum.
Therefore, Theorem \ref{them:2busmag} also holds for the case where $V_2$ is inactive and $V_1$ is binding.

\end{document}